\documentclass[twocolumn]{aastex63}

\usepackage[fleqn]{amsmath}
\usepackage[hyphenbreaks]{breakurl}
\makeatletter
\newcommand*{\rom}[1]{\expandafter\@slowromancap\romannumeral #1@}
\makeatother

\received{31 July 2020}
\revised{\today}
\accepted{\today}
\submitjournal{AJ}

\shorttitle{$JHK_s$ observations of RR Lyrae stars in M3}
\shortauthors{Bhardwaj A. et al.}

\begin{document}
\title{Near-infrared Census of RR Lyrae variables in the Messier 3 globular cluster and the Period--Luminosity Relations}

\correspondingauthor{Anupam Bhardwaj}
\email{anupam.bhardwajj@gmail.com; abhardwaj@pku.edu.cn}
\author[0000-0001-6147-3360]{Anupam Bhardwaj}\thanks{IAU Gruber Foundation Fellow 2020}
\affil{Kavli Institute for Astronomy and Astrophysics, Peking University, Yi He Yuan Lu 5, Hai Dian District, Beijing 100871, China}
\author[0000-0002-6577-2787]{Marina Rejkuba}
\affiliation{European Southern Observatory, Karl-Schwarzschild-Stra\ss e 2, 85748, Garching, Germany}
\author[0000-0002-7203-5996]{Richard de Grijs}
\affiliation{Department of Physics and Astronomy, Macquarie University, Balaclava Road, Sydney, NSW 2109, Australia}
\affiliation{Research Centre for Astronomy, Astrophysics and Astrophotonics, Macquarie University, Balaclava Road, Sydney, NSW 2109, Australia}
\affiliation{International Space Science Institute -- Beijing, 1 Nanertiao, Zhongguancun, Hai Dian District, Beijing 100190, China}
\author[0000-0001-6147-3360]{Gregory J. Herczeg}
\affil{Kavli Institute for Astronomy and Astrophysics, Peking University, Yi He Yuan Lu 5, Hai Dian District, Beijing 100871, China}
\author[0000-0001-6802-6539]{Harinder P. Singh}
\affiliation{Department of Physics and Astrophysics, University of Delhi,  Delhi-110007, India }
\author{Shashi Kanbur}
\affiliation{State University of New York, Oswego, NY 13126, USA}
\author[0000-0001-8771-7554]{Chow-Choong Ngeow}
\affil{Graduate Institute of Astronomy, National Central University, 300 Jhongda Road, 32001 Jhongli, Taiwan}


\begin{abstract} 
We present new near-infrared ($JHK_s$) time-series observations of RR Lyrae variables in the Messier 3 (NGC 5272) globular cluster using the WIRCam instrument at the 3.6-m
Canada France Hawaii Telescope. Our observations cover a sky area of $\sim 21'\times 21'$ around the cluster center and provide an average of twenty epochs of homogeneous 
$JHK_s$-band photometry. New homogeneous photometry is used to estimate robust mean magnitudes for 175 fundamental-mode (RRab), 47 overtone-mode (RRc), and 11 mixed-mode (RRd) variables. Our sample 
of 233 RR Lyrae variables is the largest thus far obtained in a single cluster with time-resolved, multi-band near-infrared photometry. 
Near-infrared to optical amplitude ratios for RR Lyrae in Messier 3 exhibit a systematic increase moving from RRc to short-period ($P < 0.6$~days) and long-period ($P \gtrsim 0.6$~days) RRab variables. 
We derive $JHK_s$-band Period--Luminosity relations for RRab, RRc, and the combined sample of variables. Absolute calibrations based on the theoretically predicted Period--Luminosity--Metallicity relations for RR Lyrae stars yield a distance modulus, $\mu = 15.041 \pm 0.017~(\textrm{statistical}) \pm 0.036~(\textrm{systematic})$~mag, to Messier 3. When anchored to trigonometric parallaxes for nearby RR Lyrae stars from the {\it Hubble Space Telescope} and the {\it Gaia} mission, our distance estimates are consistent with those resulting from the theoretical calibrations, albeit with relatively larger systematic uncertainties. \\
\end{abstract}

\section{Introduction}

RR Lyrae (RRL) variables are low-mass ($0.5\lesssim {M/M}_\odot \lesssim 0.8$), old ($>10$ Gyr) stars that are located in a region between the cross-section of the horizontal branch and the classical ``instability strip'' in the Hertzsprung--Russell diagram. These horizontal branch stars pulsate during their central helium burning evolutionary phase, similar to intermediate-mass ($3\lesssim {M/M}_\odot \lesssim 10$) classical Cepheids. RRL follow a visual ($V$-band) magnitude--metallicity relation with negligible dependence on pulsation periods unlike classical Cepheids \citep{bono2003}. The reason for this different behavior is that the bolometric correction's sensitivity to effective temperature becomes significant only at longer wavelengths \citep[$R$-band onwards,][]{catelan2004}. Indeed, RRL exhibit well defined Period--Luminosity relations (PLRs) at infrared wavelengths, first demonstrated in pioneering work by \citet{longmore1986}, which makes them excellent distance indicators \citep[see recent reviews,][]{beaton2018, bhardwaj2020}. RRL play a key role in our understanding of stellar evolution and pulsation \citep{catelan2009}, and as stellar population tracers for Galactic archaeology and the cosmic distance scale \citep{kunder2018, beaton2018}. 
  
Globular clusters (GCs) typically host a rich and homogeneous population of RRL stars. Messier 3 (M3 or NGC 5272), located at a distance of $\sim10$~kpc, hosts one of the largest samples of RRL with a dominant population of fundamental mode RRL (RRab) variables \citep{clement2001}. M3 has a mean metallicity of [Fe/H]$\sim-1.5$~dex \citep{harris2010} and the observed period distribution of its RRL population exhibits a sharp peak at fundamental pulsator period of 0.55 day \citep{jurcsik2017}, indicating that it is a typical Oosterhoff I type \citep[OoI,][]{oosterhoff1939, fabrizio2019} cluster. While multiple stellar populations have been detected along the red giant branch of M3 \citep[e.g.,][]{massari2016, lee2020}, no significant variation has been detected in the iron abundance \citep{sneden2004}. Furthermore, helium enhancement ($\Delta Y\lesssim0.02$) has been suggested to explain observed properties of horizontal branch stars \citep{dalessandro2013, valcarce2016, denissenkov2017}.

Insignificant interstellar reddening \citep{vandenberg2016} and the close proximity of M3 motivated several detailed long-term photometric investigations at optical wavelengths \citep[][and references therein]{bakos2000, cacciari2005, benko2006, jurcsik2012, jurcsik2017}. Optical photometry has been used to investigate the Blazhko effect, multimode pulsations and period doubling in M3 RRL variables \citep{jurcsik2015, jurcsik2019}. The RRL population in M3 cluster has also been explored at ultraviolet wavelengths \citep{siegel2015}. At near-infrared (NIR) wavelengths, \citet{longmore1990} derived $K_s$-band PLRs for RRL in GCs including 49 variables in the outer region of M3. Apart from that, NIR photometry of RRL in M3 has been limited to a sample of 7 RRL in the inner region of the cluster \citep{butler2003}.

M3 has been the subject of several theoretical studies aimed at reproducing the observed pulsation properties and, in particular, the period distribution of its RRL population \citep{catelan2004a, castellani2005, fadeyev2019}. \citet{catelan2004a} showed that the predicted period distribution based on canonical horizontal branch models is inconsistent with observations, while \citet{castellani2005} suggested that a bimodal mass distribution would be required to reproduce the period distribution with canonical models. \citet{marconi2007a} accurately modeled optical light curves of M3 RRL using nonlinear pulsation models with [Fe/H]$\sim-1.34$~dex \citep{carretta1997}, and estimated a distance modulus of $15.10\pm0.10$ mag. Using horizontal branch models, \citet{denissenkov2017} found a good agreement with observed properties of RRL and non-variable horizontal branch stars for a distance modulus and reddening of $\mu = 15.02$~mag and $E(B-V)=0.013$~mag, respectively.

RRL as distance indicators have gained significance with increasing NIR observations especially in GCs \citep{sollima2006, coppola2011, stetson2014, braga2015, navarrete2015, braga2018}. These horizontal branch variables can complement the tip of the red giant branch stars to provide an absolute primary calibration for the population II distance ladder \citep{beaton2016}. The homogeneous population of RRL in different GCs offers the possibility to derive their PLRs and estimate the dependence on metal abundance \citep{sollima2006}. While theoretical models predict a significant metallicity coefficient of the RRL Period--Luminosity--Metallicity (PLZ) relation \citep[e.g.,][]{catelan2004, marconi2015}, it is still a topic of active debate considering the paucity of RRL with both high-resolution spectroscopic metallicities and precise parallaxes suitable to establish an empirical calibration \citep{muraveva2018, neeley2019, bhardwaj2020}. Therefore, NIR observations of abundant RRL in M3 will not only be useful for studies of the distance scale but also complement optical and ultraviolet data for a rigorous comparison with evolutionary and pulsation models. 
  
In this work, we present NIR time-series observations of RRL in M3 for the largest sample of variables in an individual GC. The paper is organized as follows. In Section~\ref{sec:data}, we describe the observations, the data reduction and the photometric calibrations. The NIR light curves and pulsation properties of the RRL are discussed in Section~\ref{sec:rrl_phot}. We discuss the $JHK_s$-band PLRs for M3 RRL in Section~\ref{sec:rrl_plr} and estimate a robust distance to the cluster. The results are summarized in Section~\ref{sec:discuss}.

\section{Observations, data reduction, and photometric calibration} \label{sec:data}

\begin{deluxetable*}{cccccccccccccc}
\tablecaption{Log of NIR observations. \label{tbl:data}}
\tabletypesize{\footnotesize}
\tablewidth{0pt}
\tablehead{\colhead{} &  \multicolumn{4}{c}{$J$-band}  &  \multicolumn{4}{c}{$H$-band}  &  \multicolumn{4}{c}{$K_s$-band}   & \\
\colhead{Date} & \colhead{MJD}	& \colhead{Airmass} & \colhead{IQ} & \colhead{$N_\textrm{f}$} & \colhead{MJD}& \colhead{Airmass} & \colhead{IQ} & \colhead{$N_\textrm{f}$} & \colhead{MJD}& \colhead{Airmass} & \colhead{IQ} & \colhead{$N_\textrm{f}$} & ET\\
		&days	&	& arcsec	&	&days	&	& arcsec	&	&days	&	& arcsec	& 	&sec}
\startdata
2019-05-26&   58629.2882&  1.037& 0.60& 16&  58629.2953&  1.028& 0.56& 15&  58629.3002&  1.023& 0.51& 15 &5\\
2019-05-26&   58629.2903&  1.034& 0.59& 15 &   ---  &    --- &    --- &    --- &    --- &    --- &    --- &    ---   &5\\
2019-05-26&   58629.3412&  1.015& 0.78& 15&  58629.3459&  1.017& 0.67& 15&  58629.3515&  1.021& 0.76& 15 &5\\
2019-05-26&   58629.3829&  1.066& 0.79& 15&  58629.3926&  1.088& 0.65& 16&  58629.3979&  1.102& 0.60& 15 &5\\
2019-05-26&   58629.3873&  1.076& 0.72& 15 &   ---  &    --- &    --- &    --- &    --- &    --- &    --- &    ---   &5\\
2019-05-26&   58629.4659&  1.446& 1.09& 14&  58629.4733&  1.513& 0.93& 15&  58629.4783&  1.563& 1.02& 14 &5\\
2019-05-27&   58630.2506&  1.109& 0.53& 15&  58630.2558&  1.095& 0.53& 15&  58630.2612&  1.081& 0.49& 15 &5\\
2019-05-27&   58630.2925&  1.028& 0.46& 15&  58630.2975&  1.023& 0.44& 15&  58630.3022&  1.019& 0.45& 15 &5\\
2019-05-27&   58630.3369&  1.014& 0.71& 15&  58630.3414&  1.016& 0.72& 16&  58630.3478&  1.021& 0.67& 16 &5\\
2019-05-27&   58630.3783&  1.062& 0.77& 15&  58630.3831&  1.072& 0.65& 15&  58630.3879&  1.083& 0.61& 15 &5\\
2019-05-27&   58630.4223&  1.197& 0.59& 15&  58630.4269&  1.217& 0.57& 15&  58630.4314&  1.239& 0.49& 15 &5\\
2019-05-27&   58630.4666&  1.476& 0.65& 15&  58630.4711&  1.519& 0.60& 15&  58630.4757&  1.563& 0.53& 15 &5\\
2019-05-28&   58631.2576&  1.083& 0.72& 15&  58631.2627&  1.072& 0.68& 15&  58631.2672&  1.062& 0.63& 15 &5\\
2019-05-28&   58631.3046&  1.016& 0.70& 15&  58631.3100&  1.013& 0.70& 15&  58631.3151&  1.012& 0.73& 15 &5\\
2019-05-28&   58631.3475&  1.023& 0.74& 15&  58631.3522&  1.027& 0.72& 15&  58631.3579&  1.034& 0.73& 15 &5\\
2019-05-28&   58631.3887&  1.092& 0.64& 15&  58631.3933&  1.104& 0.66& 15&  58631.3979&  1.118& 0.58& 15 &5\\
2019-05-28&   58631.4323&  1.258& 0.71& 15&  58631.4369&  1.283& 0.66& 15&  58631.4416&  1.311& 0.67& 15 &5\\
2019-05-28&   58631.4771&  1.608& 0.83& 15&  58631.4817&  1.662& 0.84& 15&  58631.4863&  1.721& 0.79& 15 &5\\
2019-05-29&   58632.3011&  1.018& 0.89& 16&  58632.3104&  1.012& 0.84& 15&  58632.3160&  1.011& 0.67& 15 &5\\
2019-05-29&   58632.3587&  1.039& 0.75& 15&  58632.3634&  1.046& 0.77& 15&  58632.3680&  1.053& 0.68& 15 &5\\
2019-05-29&   58632.4045&  1.148& 0.60& 15&  58632.4093&  1.166& 0.62& 15&  58632.4141&  1.185& 0.56& 15 &5\\
2019-05-29&   58632.4509&  1.392& 0.60& 15&  58632.4556&  1.429& 0.63& 15&  58632.4605&  1.470& 0.56& 15 &5\\
\enddata
\tablecomments{MJD: Modified Julian Date (JD$-$2,400,000.5). IQ: Image quality (in arcseconds) measured by the queued service observing at the CFHT. $N_\textrm{f}$: Number of dithered frames per epoch. 
ET: Exposure time (in seconds) for each dithered frame.}
\end{deluxetable*}

\subsection{Observations and data reduction}

Our NIR observations were obtained using the WIRCam instrument \citep{puget2004} mounted on the 3.6-m Canada France Hawaii Telescope (CFHT) on the summit of Mauna Kea in Hawaii during 
4 nights between 26 and 29 May 2019. WIRcam is an array of four $2048\times2048$ HgCdTe HAWAII-RG2 detectors arranged in $2\times2$ grid with gaps of $45\arcsec$ between adjacent detectors. The pixel scale of each detector is $0.3\arcsec$ pixel$^{-1}$ resulting in a field of view of $\sim21'\times 21'$. We requested $JHK_s$ time-series observations in queue mode centered on the M3 cluster center, and obtained 22 epochs in $J$ and 20 epochs in the $H$ and $K_s$-bands. Each epoch consisted of on average 15 dithered images obtained with an exposure time of 
5s per image. This resulted in more than 900 images in total. A summary of all the epochs in $JHK_s$-bands is listed in Table~\ref{tbl:data}.  

Images were downloaded from the \texttt{IDL} Interpretor of the WIRCam Images (\texttt{$`$I$`$iwi}{\footnote{\url{https://www.cfht.hawaii.edu/Instruments/Imaging/WIRCam/IiwiVersion2Doc.html}}}) preprocessing pipeline at CFHT. The \texttt{$`$I$`$iwi} pipeline incorporates detrending (dark subtraction, flat-fielding) and initial sky subtraction, and provides calibrated WIRCam data products.
For each preprocessed image, a weight map was created using \texttt{WeightWatcher} \citep{marmo2008} to mask bad pixels in the WIRCam mosaic. Astrometric calibration of preprocessed images was performed using \texttt{SCAMP} \citep{bertin2006}. \texttt{SCAMP} uses a catalog of sources matched with the Two Micron All Sky Survey (2MASS) Point Source Catalog \citep{skrutskie2006} generated using \texttt{SExtractor} \citep{bertin1996}. The astrometric calibration was done at a very high precision both internally ($\sim\sigma_{\textrm{int}}=0.1\arcsec$) and externally using 2MASS ($\sim\sigma_{\textrm{ext}}=0.15\arcsec$). \texttt{SCAMP} also scales the flux of each detector with different magnitude zero-points and performs an initial photometric calibration  against 2MASS with a root-mean-square ({\it rms}) error of $\sim 0.025$ mag for high signal-to-noise ($S/N > 100$), and with a {\it rms} of $\sim 0.045$~mag for fainter stars. After performing astrometric calibration, dithered images at each epoch were median-combined using \texttt{SWARP} \citep{bertin2002} at the instrument pixel scale. 

\begin{figure}
\epsscale{1.2}
\plotone{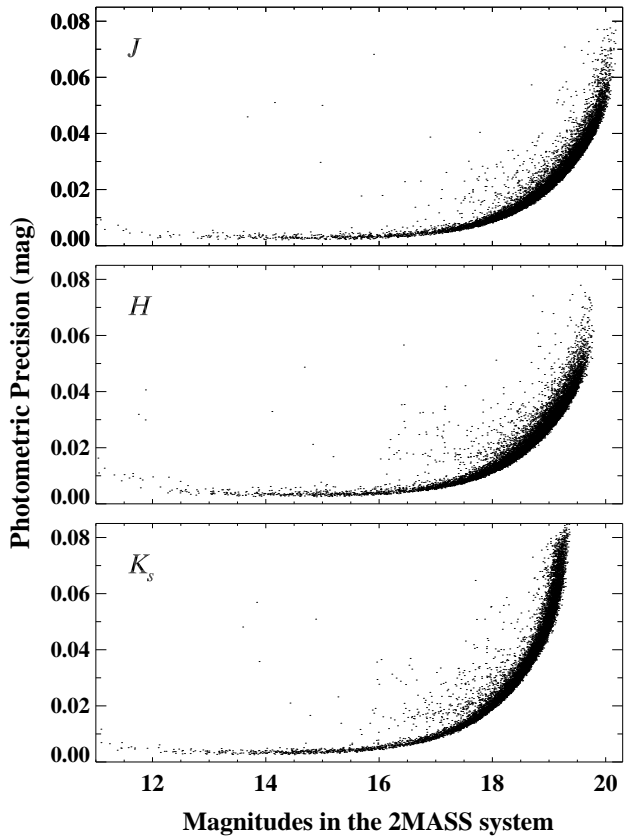}
\caption{Internal photometric precision of our photometry as a function of 2MASS magnitude for the $J$ (top), $H$ (middle), and $K_s$ (bottom) bands. In all three panels, we have excluded sources (1) with $\sigma_\textrm{external}/\sigma_\textrm{internal} \lesssim 2$ in all three bands; (2) that are located within 300 pixels in radius from the crowded center; (3) that are within 300 pixels from the corners of the detectors.} 
\label{fig:sig_phot}
\end{figure}

\subsection{Point-spread function photometry}

We performed photometry on each epoch image using the \texttt{DAOPHOT/ALLSTAR} \citep{stetson1987} and \texttt{ALLFRAME} \citep{stetson1994} routines applied to the $J$, $H$, and $K_s$ filters separately. As a first step, we determined an approximate full width at half maximum (FWHM) for sources in each image using \texttt{IRAF}{\footnote{\texttt{IRAF} is distributed by the National Optical Astronomy Observatory, which is operated by the Association of Universities for Research in Astronomy (AURA) under cooperative agreement with the National Science Foundation.}. Using \texttt{DAOPHOT}, we identified all sources $>4\sigma$ detection threshold and performed aperture photometry within 3 pixel apertures. In the second step, we selected up to 300 bright and isolated stars uniformly distributed across each image excluding sources in the inner 500 pixels from the crowded center of the cluster. These stars were selected to determine a point-spread function (PSF) for each image. The PSF was modeled as a Gaussian profile with no spatial variation across the detector. PSF photometry was performed using \texttt{ALLSTAR} on all sources for which aperture photometry was obtained in the first step. Finally, accurate frame-to-frame coordinate transformations were obtained for all epoch images using the \texttt{DAOMATCH} and \texttt{DAOMASTER} routines \citep{stetson1993}. 

In the third step, we combined best-seeing (IQ $<0.5$) $JHK_s$-band epoch images based on the FWHM to create a higher $S/N$ reference frame. The first two steps were repeated to obtain a common star list for each filter. Similarly, frame-to-frame coordinate transformations were also derived with respect to the reference frame for all epoch images. The reference star list was used as input for the PSF photometry in the \texttt{ALLFRAME} routine. Output photometry at each epoch was merged to obtain light curves and mean magnitudes in each filter for sources that were observed in at least 10 epochs. We also used Stetson's \texttt{TRIAL} program to extract light curves of candidate variables and determine mean instrumental magnitudes and variability indices \citep{stetson1996}. The internal photometric precision of the $JHK_s$ magnitudes is shown as a function of 2MASS magnitude in Fig.~\ref{fig:sig_phot} after excluding sources in the most crowded central region of the cluster.

\subsection{Photometric calibration in the 2MASS system}

The photometric catalogs in the $J$, $H$ and $K_s$ filters were matched and merged using \texttt{DAOMATCH} and \texttt{DAOMASTER} to perform the final photometric calibration. We found 1968 2MASS stars in our field of view and restricted the sample to stars with photometric quality flag `AAA'. This flag implies that the photometric measurements in all three $JHK_s$-bands are determined with a $S/N\gtrsim 10$. Sources located within $2'$ from the crowded center were also excluded to avoid blended objects. Furthermore, the sample was limited to objects with 2MASS magnitudes fainter than 11~mag in the $JHK_s$-bands to avoid saturation and nonlinearities. After these restrictions, we cross-matched the merged catalog with the 2MASS stars and found 552 stars in common within a tolerance of $1''$. 

For absolute photometric calibration, we first corrected for a fixed magnitude-independent zero-point offset between 2MASS and instrumental magnitudes in the $JHK_s$-bands. Next, we solved for a color dependence by employing linear color terms in the transformations. Individual objects with residuals greater than $3\sigma$ from the initial fits were discarded iteratively to obtain robust transformations ({\it rms} of $\sim0.05$~mag for each fit in the $JHK_s$-bands). We found a statistical dependence on the 2MASS color term but adding this extra parameter did not contribute to any significant reduction in the {\it rms} or the chi-squared per degree of freedom. Note that the majority of 2MASS standards span a relatively narrow range in color ($\Delta(J-K_s) \lesssim$ 0.8~mag, $\Delta(H-K_s) \lesssim$ 0.4~mag). Furthermore, the uncertainties in the 2MASS colors are significant (up to $\sim 0.15$ mag for quality flag `A') while the uncertainties in the instrumental magnitudes are $5-10\times$ smaller. We also derived transformations including instrumental color terms. No significant dependence on instrumental color term was found and therefore, we did not apply any color corrections. We estimated a maximum uncertainty of $\lesssim 0.03$~mag in the photometry corresponding to the color range of RRL stars in common with 2MASS in our photometric catalogs.

\begin{figure}
\epsscale{1.2}
\plotone{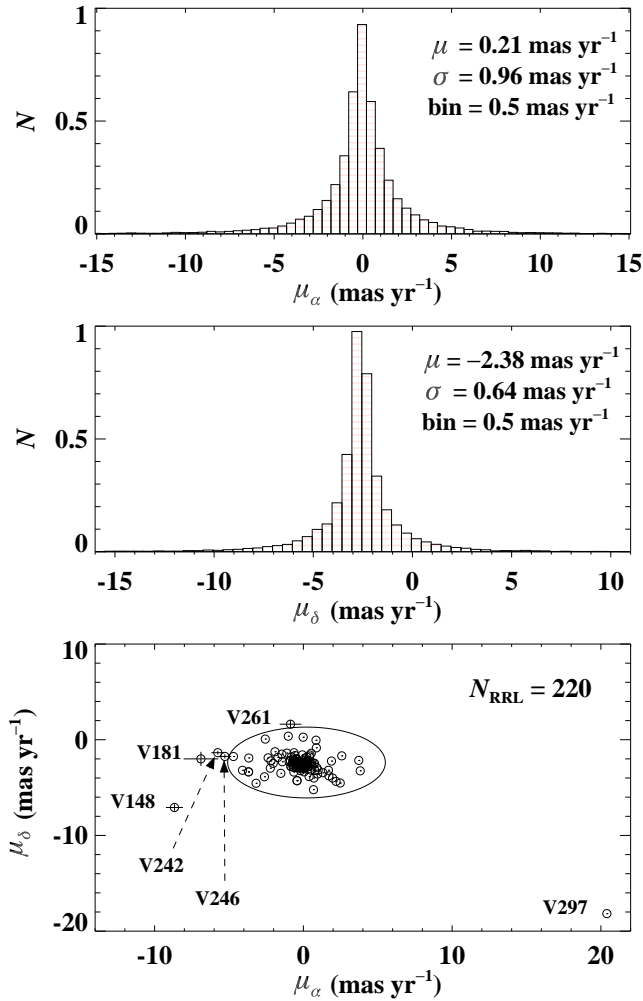}
\caption{Histograms of proper motions along right ascension ({top}) and declination ({middle}) for {\it Gaia} sources in the WIRCam field of view. The mean values and the standard deviations of the Gaussian fits to the histograms are also shown in the top and middle panels. {Bottom:} Scatter plot of proper motions of RRL variables in the M3 cluster (see Section~\ref{sec:rrl_phot}) for which {\it Gaia} astrometry is available. Median error bars are of the order of the symbol size. An ellipse corresponding to $\pm5\sigma$ standard deviations about the mean proper motions is shown and the outliers beyond this threshold are also tagged with their IDs.} 
\label{fig:pms}
\end{figure}

\subsection{M3 photometry and proper motions}

We cross-matched our NIR photometric catalog with the second data release from the {\it Gaia} mission \citep[DR2,][]{lindegren2018}, and found 27,417 objects for which proper motions and $G$-band photometry are available. The matching radius was set to $1\arcsec$ and the nearest neighbor was adopted in case more than one was found within this radius. The top and middle panels of Fig.~\ref{fig:pms} display the histograms of proper motions of stars within the WIRCam field of view. The histograms of proper motions along the right ascension and declination peak at $\mu_\alpha = 0.211$ and $\mu_\delta = -2.385$ mas yr$^{-1}$ with a half width at half maximum of $1.129$ and $0.752$~mas yr$^{-1}$, respectively. The mean proper motions are consistent with those derived by 
\citet[$\mu_\alpha = -0.11$, $\mu_\delta = -2.63$,][]{helmi2018} considering the large standard deviation of the Gaussian distribution. Given the uncertainties in the astrometry, we conservatively consider all sources within $\pm 5\sigma$ of their peak proper motions as members of the cluster. Fig.~\ref{fig:cmd_all} displays the proper motion cleaned $(J-K_s), K_s$ color--magnitude diagram for sources in M3. The proper motions of RRL are shown in the bottom panel of Fig.~\ref{fig:pms}. The location of RRL on the horizontal branch is also shown in Fig.~\ref{fig:cmd_all} using mean magnitudes determined in the next section.

\begin{figure}
\epsscale{1.2}
\plotone{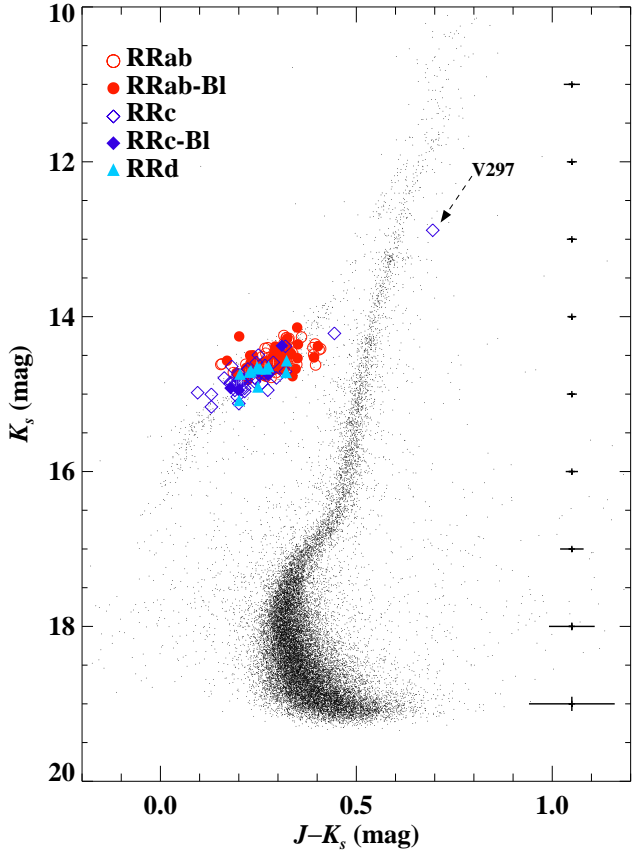}
\caption{Color--magnitude diagram for stars in M3 with three-band photometry, and for which the proper motions are consistent within $\pm 5\sigma$ of their mean values. Candidate RRL variables (see Section~\ref{sec:rrl_phot}) are overplotted. RRab-Bl and RRc-Bl represent RRL stars known to display the Blazhko effect. Representative $\pm2\sigma$ error bars in both the magnitudes and colors are also shown.} 
\label{fig:cmd_all}
\end{figure}

\section{RR Lyrae Photometry}
\label{sec:rrl_phot}

We adopted a reference list of variable candidates in M3 from the updated catalog{\footnote{\url{http://www.astro.utoronto.ca/~cclement/}}} of \citet{clement2001}. Their compilation consists of 241 RRL stars including coordinates, periods, $V$-band amplitudes{\footnote{We exclusively use $A_\lambda$ to refer to the amplitudes in a given filter and not the extinction corrections.}}, and the classification for most of these cluster variables. There are 178 RRab, 48 overtone-mode RRL (RRc), and 11 double/multi-mode (RRd) variables. Four RRL (V129, V217, V265, and V268) have uncertain classifications and two of these (V265 and V268) do not have any determination of their pulsation period. Six of the 241 RRL variables (V113, V115, V123, V205, V206, V299) are outside the WIRCam field of view. The periods, Oosterhoff and Blazhko types for these variables were updated following \citet{jurcsik2015} and \citet{jurcsik2017}. 

The $JHK_s$ light curves of the RRL were extracted using a cross-match with PSF photometric catalogs within a search radius of $1\arcsec$. While $90\%$ of targets matched within $0.1\arcsec$ tolerance, photometry for two RRL (V191 and V192){\footnote{V191 and V192 are located in the unresolved central $1.5'$ of the cluster and their photometry is also contaminated.}} was retrieved with $\sim 1.2\arcsec$. We also computed periods for the well-sampled light curves and found good agreement with periods compiled by \citet{clement2001}. The latter periods were used to phase the light curves of all variables. Note that all mixed mode variables were phased with their dominant first-overtone periods. We also determined a period of 0.5284 days for V265 which has no period listed in the catalog of \citet{clement2001}. However, our photometry of the significantly blended V268 did not allow a period determination for this variable, and therefore, it is excluded from our analysis. The final sample of RRL includes 234 stars (175 RRab, 48 RRc, 11 RRd).

\begin{deluxetable}{rrrrrr}
\tablecaption{NIR time-series photometry of RRL in the M3 cluster. \label{tbl:phot_rrl}}
\tabletypesize{\footnotesize}
\tablewidth{0pt}
\tablehead{\colhead{ID} & \colhead{Band} & \colhead{MJD} & \colhead{Mag.} & \colhead{$\sigma_{\textrm{mag}}$}& \colhead{QF}}
\startdata
   V1&   $J$ &   58629.2882 &    14.653 &     0.021 &  A \\
   V1&   $J$ &   58629.3412 &    14.736 &     0.022 &  A \\
   V1&   $J$ &   58629.3829 &    14.748 &     0.010 &  A \\
     ... &      ... &      ... &      ... &      ... &         \\
   V1&   $H$ &   58629.2953 &    14.526 &     0.020 &  A \\
   V1&   $H$ &   58629.3459 &    14.511 &     0.019 &  A \\
   V1&   $H$ &   58629.3926 &    14.589 &     0.019 &  A \\
     ... &      ... &      ... &      ... &      ... &         \\
   V1& $K_s$ &   58629.3002 &    14.502 &     0.023 &  A \\
   V1& $K_s$ &   58629.3515 &    14.526 &     0.014 &  A \\
   V1& $K_s$ &   58629.3979 &    14.456 &     0.014 &  A \\
     ... &      ... &      ... &      ... &      ... &         \\
\enddata
\tablecomments{ID: Same as in the catalog of \citet{clement2001}; MJD $=$ JD $- 2,400,000.5$. The fourth column represents magnitude in a NIR band, and the fifth column lists its associated
uncertainty. QF - Quality flag. This table is available in its entirety in machine-readable form. A sample time-series in $JHK_s$ for a RRL is shown here for guidance regarding its content.}
\end{deluxetable}

\begin{figure*}
\epsscale{1.2}
\plotone{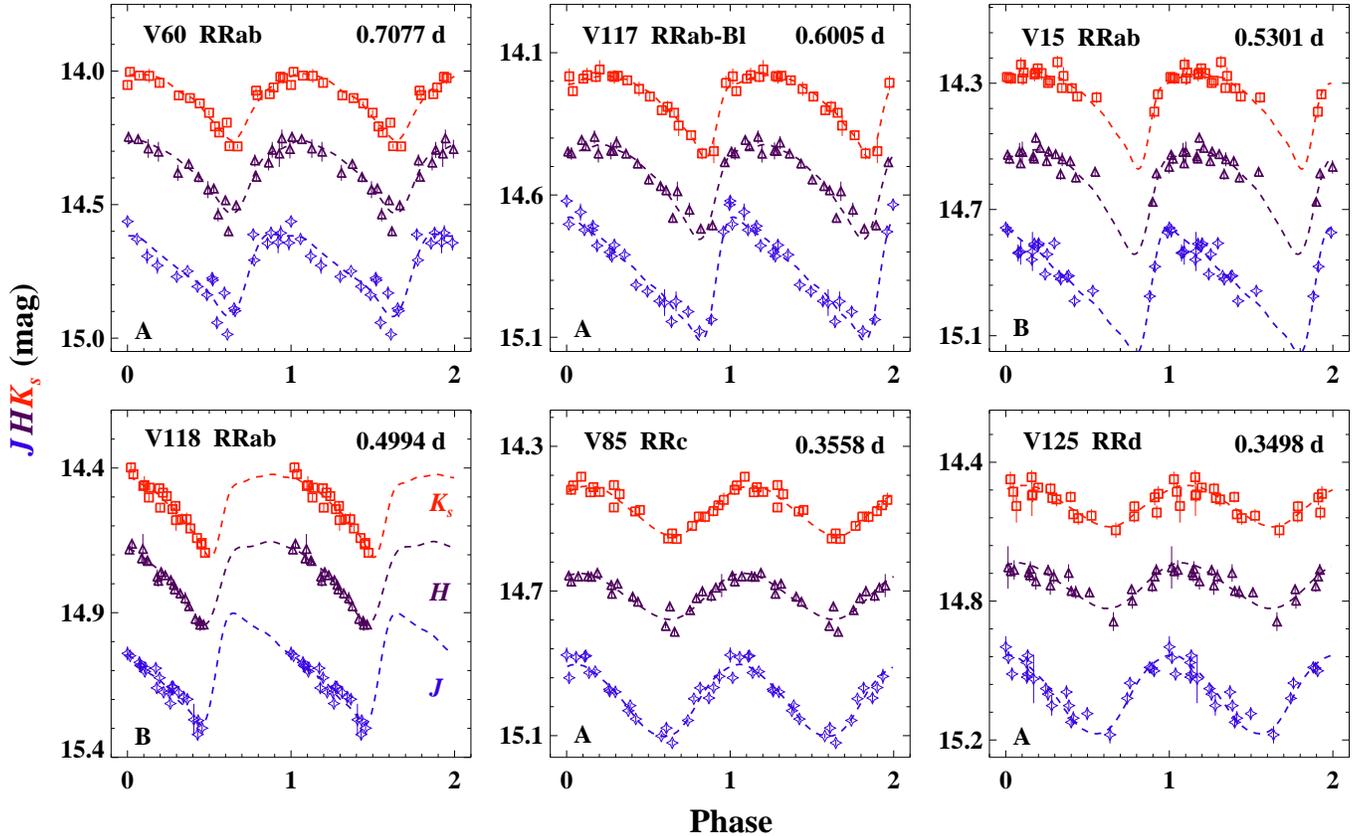}
\caption{Representative $JHK_s$-band light curves of different subclasses of RRL spanning the entire range of periods in our sample. The $J$ (blue stars) and $K_s$ (red squares) 
light curves are offset for clarity by $+0.1$ and $-0.2$ mag, respectively. The dashed lines represent the best-fitting templates to the data in each band. The mixed-mode variable (V125) is phased according to its first-overtone period. Star ID, subtype, and the pulsation period are included at the top of each panel. Light curve quality flags are also included at the bottom left of
each panel.}
\label{fig:lcs}
\end{figure*}

The light curves were fitted using a fourth-order Fourier sine series \citep[e.g.,][]{bhardwaj2015} to inspect their quality and determine phase differences ($\Delta \phi$) between successive observations. Initially, light curves with a maximum of $\Delta \phi \lesssim 0.2$ and {\it rms} $\lesssim 0.05$~mag with respect to the Fourier fits were assigned `A' quality flags while the remaining light curves were flagged as `B'.  However, most RRL with periods $0.47 < P < 0.53$ days exhibit larger phase gaps ($\Delta \phi > 0.2$) either around mean-light or near the extrema. Therefore, Fourier-fitted light curves were also inspected visually and flagged as `A' if the extrema were well-constrained so as to estimate accurate amplitudes. The poor-quality light curves which exhibit large scatter or do not show any distinct periodicity in one or more filters, due to photometric contamination, were assigned a `C' quality flag. Fig.~\ref{fig:lcs} displays a few example light curves of  quality flags `A' and `B', and different subclasses of RRL stars spanning the entire period range (see also Appendix~\ref{sec:app_b}). NIR time-series photometry of M3 RRL is provided in Table~\ref{tbl:phot_rrl}.

\subsection{Template-fits, amplitude ratios and mean magnitudes}

\begin{figure}
\epsscale{1.2}
\plotone{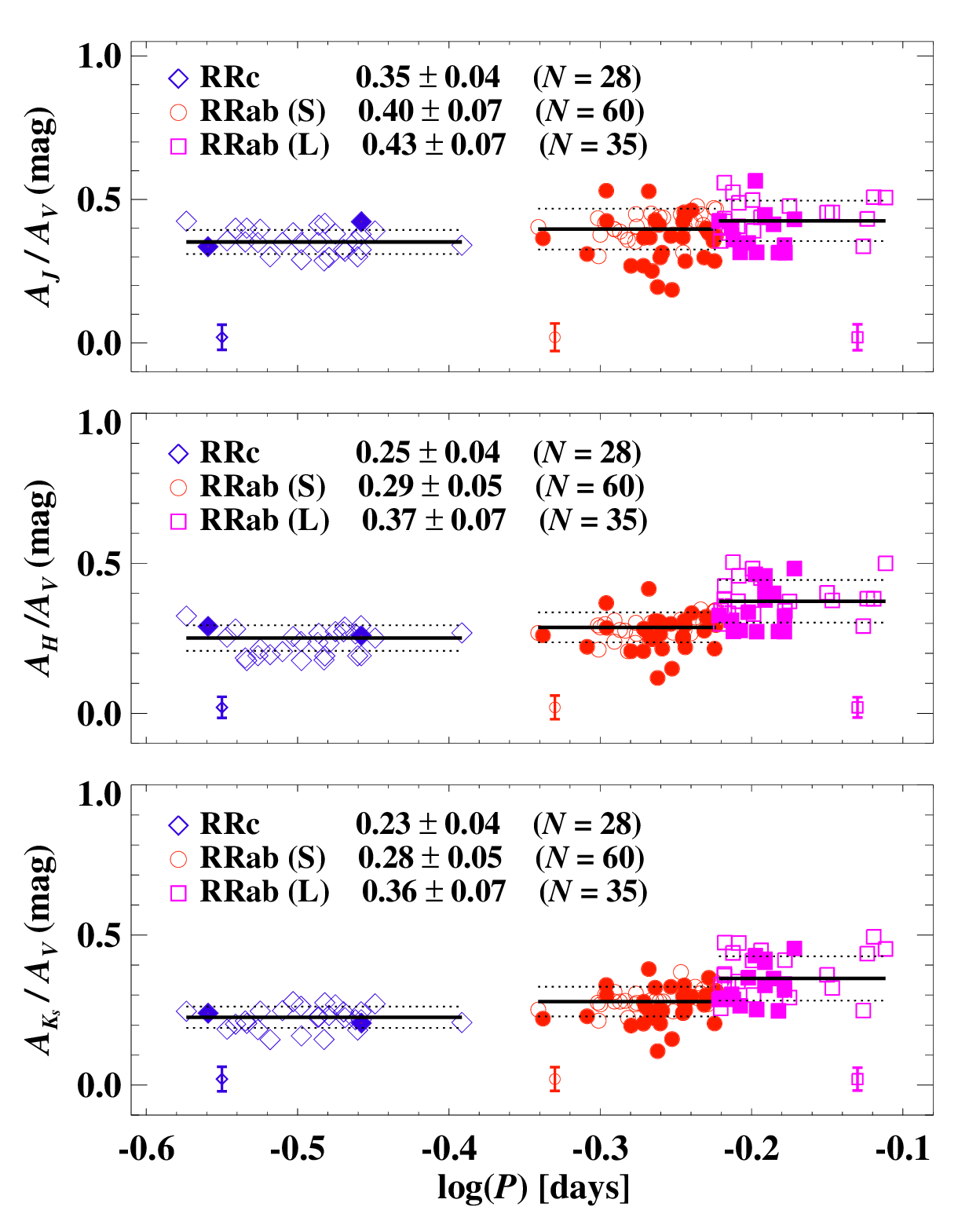}
\caption{NIR-to-optical amplitude ratios - $A_J/A_V$ (top), $A_H/A_V$ (middle), and $A_{K_s}/A_V$ (bottom) are plotted as a function of the logarithmic period. The median value and the standard deviation ($M \pm \sigma$), and the number of stars for each sample of RRc, short-period ($P \lesssim 0.6$~day) RRab, and long-period ($P > 0.6$~day) RRab are also shown in each panel. The solid and dashed lines represent the median and $\pm1\sigma$ standard deviation of each sample. RRL stars known to display the Blazhko effect are shown using filled symbols. Representative median error bars are also shown at the bottom of each panel.} 
\label{fig:nir_av}
\end{figure}
 
\begin{figure}
\epsscale{1.2}
\plotone{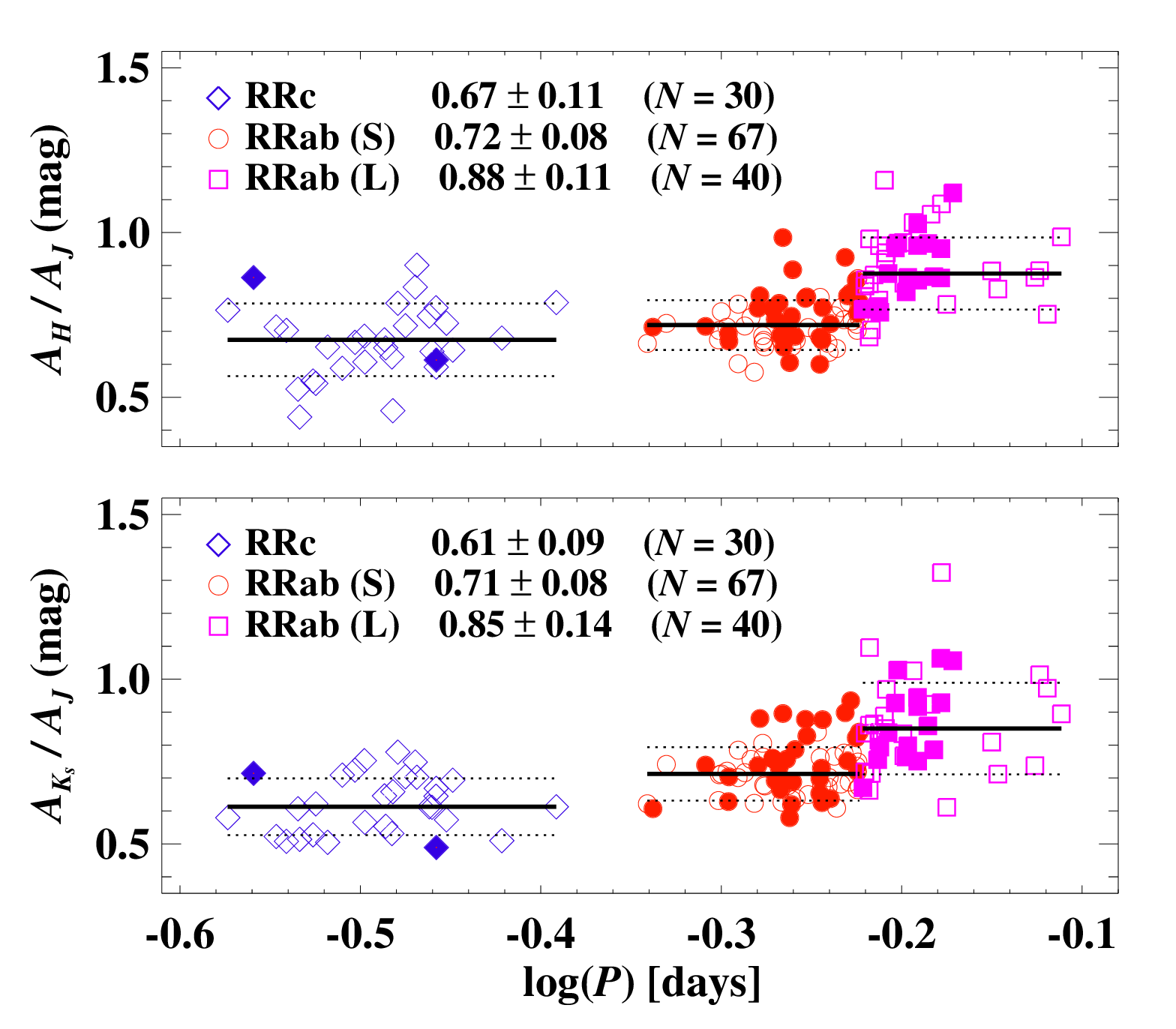}
\caption{As Fig.~\ref{fig:nir_av} but for the NIR amplitude ratios - $A_H/A_J$ (top) and $A_{K_s}/A_J$ (bottom). Median error bars are of the order of the symbol size.} 
\label{fig:nir_amp}
\end{figure}

NIR light curve templates are useful to estimate robust mean magnitudes for RRL having sparsely sampled light curves. New NIR templates for RRab and RRc stars were provided by \citet{braga2019} covering
three period bins ($P\lesssim0.55$, $ 0.55 < P < 0.7$, and $P \gtrsim 0.7$~day) for RRab and a single period bin for all RRc stars. Initially, we fitted templates to RRL light 
curves with quality flag `A' solving for a phase offset and amplitude simultaneously. The peak-to-peak amplitudes were determined accurately with a median uncertainty of 33, 28, and
30 mmag in the $J$, $H$, and $K_s$-bands, respectively. These amplitude measurements are critical to constrain the amplitudes for the light curves having large phase gaps when combined with the known optical amplitudes, and to determine mean magnitudes.

Fig.~\ref{fig:nir_av} displays NIR-to-optical amplitude ratios for RRL with well-sampled $JHK_s$ light curves. \citet{braga2018} provided empirical evidence that NIR-to-optical amplitude ratios for the long-period ($P \gtrsim 0.7$~day) RRab in $\omega$ Cen are systematically larger than for the short-period ($P < 0.7$~day) RRab. In Fig.~\ref{fig:nir_av}, a similar trend is also seen for long-period ($P \gtrsim 0.6$~day) RRab in M3. The period at which this shift occurs is smaller for RRab in M3 than for those in $\omega$ Cen. The increase in the amplitude ratios for long-period ($P \gtrsim 0.6$~day) RRab is significant in case of $H$ and $K_s$-bands. Median values of NIR-to-optical amplitude ratios for short-period ($P < 0.6$~day) M3 RRab are identical to those for RRab ($P < 0.7$~day) in $\omega$ Cen in the case of $A_H/A_V$ and $A_{K_s}/A_V$. For RRc in the $J$-band and long-period ($P \gtrsim 0.6$~day) RRab, the median values are typically smaller for M3 variables compared to those of RRL in the $\omega$ Cen. Some of the Blazhko variables seem to be outliers in the amplitude ratio planes but the dichotomy feature in amplitude ratios remains even if we exclude Blazhko variables. Furthermore, we found consistent results if the amplitudes were determined directly from the time-series data without template fits, but with a greater standard deviation. The mean values and the standard deviations of these amplitude ratios are listed in Table~\ref{tbl:amp_ratio}.

\begin{deluxetable}{rcccccc}
\tablecaption{Mean amplitude ratios for RRL in M3 cluster. \label{tbl:amp_ratio}}
\tabletypesize{\footnotesize}
\tablewidth{0pt}
\tablehead{\colhead{Band} & \multicolumn{2}{c}{RRc} & \multicolumn{2}{c}{RRab (S)} & \multicolumn{2}{c}{RRab (L)} \\
			&Mean & $\sigma$ & Mean & $\sigma$ & Mean & $\sigma$\\
			&	\multicolumn{6}{c}{mag}}
\startdata
			&	\multicolumn{6}{c}{All RRL with quality flag `A'}\\
\hline
             $A_{J}/A_V$ &     0.357 &       0.042 &       0.391 &       0.071 &       0.419 &       0.071\\
            $A_{H}/A_V$ &     0.237 &       0.043 &       0.281 &       0.050 &       0.373 &       0.071\\
          $A_{K_s}/A_V$ &     0.222 &       0.035 &       0.278 &       0.049 &       0.361 &       0.074\\
            $A_{H}/A_J$ &     0.671 &       0.110 &       0.721 &       0.075 &       0.891 &       0.110\\
          $A_{K_s}/A_H$ &     0.620 &       0.086 &       0.720 &       0.081 &       0.854 &       0.139\\
\hline
			&    \multicolumn{6}{c}{Non-Blazhko RRL with quality flag `A'} \\
\hline
            $A_{J}/A_V$ &     0.355 &       0.041 &       0.412 &       0.042 &       0.439 &       0.065\\
            $A_{H}/A_V$ &     0.235 &       0.043 &       0.294 &       0.036 &       0.389 &       0.063\\
          $A_{K_s}/A_V$ &     0.222 &       0.036 &       0.282 &       0.034 &       0.379 &       0.075\\
            $A_{H}/A_J$ &     0.667 &       0.107 &       0.701 &       0.057 &       0.897 &       0.118\\
          $A_{K_s}/A_H$ &     0.621 &       0.084 &       0.704 &       0.059 &       0.844 &       0.154\\
\enddata
\tablecomments{Mean and standard deviation ($\sigma$). RRab (S) : Short-period RR Lyrae ($\log(P) < 0.6$) days. RRab (L) : Long-period RR Lyrae ($\log(P) \gtrsim 0.6$) days.}
\end{deluxetable}

\begin{deluxetable*}{rccccccccccccccl}
\tablecaption{NIR pulsation properties of RRL in the M3 cluster. \label{tbl:m3_nir}}
\tabletypesize{\footnotesize}
\tablewidth{0pt}
\tablehead{\colhead{ID} & \colhead{RA} & \colhead{Dec} & \colhead{Period} & \colhead{Type}& \multicolumn{3}{c}{Mean magnitudes}  & \multicolumn{3}{c}{$\sigma_{\textrm{mag}}$} & \multicolumn{3}{c}{Amplitudes ($A_\lambda$)}  & $\Delta''$ & \colhead{QF}\\
 	&	&		& 	    &	   & $J$    & $H$     & $K_s$ & $J$     & $H$    & $K_s$ & $J$     & $H$    & $K_s$  &        &	\\  
 	&	deg.&	deg.	& days 	    &	   & \multicolumn{3}{c}{mag}  & \multicolumn{3}{c}{mag}  & \multicolumn{3}{c}{mag}   &  arcsec      &  }
\startdata
   V1 &  205.546333 &   28.342722 & 0.52059 &  RRab & 14.880 & 14.647 & 14.594 &  0.017 &  0.019 &  0.020 &  0.419 &  0.311 &  0.317 &  0.006 &  AI\\
   V3 &  205.565458 &   28.361611 & 0.55818 &  RRab & 14.916 & 14.636 & 14.607 &  0.019 &  0.023 &  0.024 &  0.416 &  0.344 &  0.325 &  0.015 & BII, Bl\\
  V4n &  205.534125 &   28.375972 & 0.58504 &  RRab & 14.623 & 14.330 & 14.263 &  0.028 &  0.036 &  0.030 &  0.341 &  0.270 &  0.213 &  0.016 &   B\\
  V4s &  205.534250 &   28.375889 & 0.59305 &  RRab & 14.766 & 14.509 & 14.609 &  0.029 &  0.036 &  0.033 &  0.443 &  0.304 &  0.274 &  0.453 &   B\\
   V5 &  205.630375 &   28.372417 & 0.50579 &  RRab & 14.881 & 14.682 & 14.662 &  0.015 &  0.015 &  0.018 &  0.557 &  0.386 &  0.350 &  0.008 & AI, Bl\\
   V6 &  205.508667 &   28.394889 & 0.51434 &  RRab & 14.968 & 14.752 & 14.679 &  0.020 &  0.021 &  0.022 &  0.438 &  0.275 &  0.272 &  0.006 &  BI\\
   V7 &  205.546208 &   28.402833 & 0.49742 &  RRab & 15.008 & 14.761 & 14.736 &  0.020 &  0.024 &  0.022 &  0.532 &  0.325 &  0.287 &  0.008 & BI, Bl\\
   V8 &  205.522083 &   28.371889 & 0.63671 &  RRab & 14.456 & 14.300 & 14.255 &  0.023 &  0.027 &  0.024 &    --- &    --- &    --- &  0.346 & C, Bl\\
   V9 &  205.456292 &   28.320361 & 0.54155 &  RRab & 14.862 & 14.622 & 14.561 &  0.016 &  0.020 &  0.018 &  0.469 &  0.314 &  0.294 &  0.004 &  AI\\
  V10 &  205.596250 &   28.416833 & 0.56955 &  RRab & 14.824 & 14.567 & 14.515 &  0.017 &  0.017 &  0.017 &  0.437 &  0.296 &  0.319 &  0.004 & AI, Bl\\
  V11 &  205.499917 &   28.319944 & 0.50789 &  RRab & 14.874 & 14.664 & 14.604 &  0.015 &  0.015 &  0.017 &  0.518 &  0.240 &  0.219 &  0.007 &  BI\\
\enddata
\tablecomments{Star ID, coordinates (epoch J2000), periods, and subtypes are taken from \citet{clement2001}. $\Delta$ is the separation, in arcseconds, between coordinates of RRL
from \citet{clement2001} and our astrometric calibration. Quality flags (QF) - `A', `B', and `C' (see text); `I' and `II' represent Oosterhoff types I and II, respectively; `Bl' indicates Blazhko variation. Photometric pulsation properties of V297 are also included for completeness. This table is available in its entirety in machine-readable form in the online journal. A portion is shown here for guidance regarding its form and content.}
\end{deluxetable*}

Fig.~\ref{fig:nir_amp} shows NIR amplitude ratios for RRL in M3. An increase in the median value of $A_H/A_J$ and $A_{K_s}/A_J$ for long-period RRab is also evident, similar to the result of \citet{braga2018}. This feature of amplitude ratios involving NIR data is different to the behavior of optical amplitude ratio ($A_{I}/A_V$) for M3 RRab \citep[][]{jurcsik2018} which is constant over the entire period range \citep[see][for $\omega$ Cen RRab]{braga2015}. While this dichotomy is apparent for RRab in the GCs, \citet{jurcsik2018} instead provided empirical evidence of a linear increase in $A_{K_s}/A_I$ as a function of period for RRab in the Galactic bulge. The dichotomy in RRab amplitude ratios is observed in GCs of two different Oosterhoff types (M3 - OoI and $\omega$ Cen - OoII) and different metallicity distributions (significant spread in $\omega$ Cen versus negligible spread in M3). Therefore, it is unlikely that metallicity is playing an important role. However, the observed period shift in the break period ($\log(P)=-0.222$ [days] for M3 versus $\log(P)=-0.155$ [days] for $\omega$ Cen) in NIR-to-optical amplitude ratios is in excellent agreement with the offset between the mean periods of their RRab stars ($\Delta \log(P_{\textrm{RRab}}) = -0.066$~[days]). This hints that the break period in the amplitude ratios involving NIR data is also an indicator of the Oosterhoff type of the cluster. Further investigation is needed to confirm the feature in the amplitude ratios and understand the cause of the dichotomy.

Finally, NIR templates were fitted to the light curves using NIR-to-optical amplitude ratios listed in Table~\ref{tbl:amp_ratio} for M3 RRL allowing for variations within $1\sigma$ of the quoted uncertainties. We used three period bins ($P\lesssim0.54$, $ 0.54 < P < 0.6$, and $P \gtrsim 0.6$~day) for RRab in contrast to \citet{braga2018}. The choice of these adopted period cuts was based on the empirical result of the amplitude ratios and the variations in the light curve parameters of RRab in M3 at these periods \citep{jurcsik2017}. The Fourier amplitude parameter ($R_{21}$) in $V$-band starts to decrease as a function of period $\sim 0.6$~day onwards and the phase parameter ($\phi_{21}$) exhibits a sudden increase for $P > 0.54$~day \citep[see Figure 6 of][]{jurcsik2017}. The lower-order Fourier parameters contain the most characteristic information about the shape of the light curves \citep{slee1981, bhardwaj2015, bhardwaj2017, das2018}.  

The mean magnitudes were estimated through numerical integration of the best-fitting templates. While the uncertainties in the mean magnitudes from the template fits were typically $<0.01$~mag, we conservatively added the median photometric error in the individual measurements to the uncertainties in the mean magnitudes. For multi-mode variables and light curves with quality flag `C', weighted mean magnitudes were simply determined from the multi-epoch measurements. The peak-to-peak amplitudes were also determined from the template fits for RRL with quality flag `B'. The NIR pulsation properties, mean magnitudes, and amplitudes are tabulated in Table~\ref{tbl:m3_nir}.

\begin{figure}
\epsscale{1.2}
\plotone{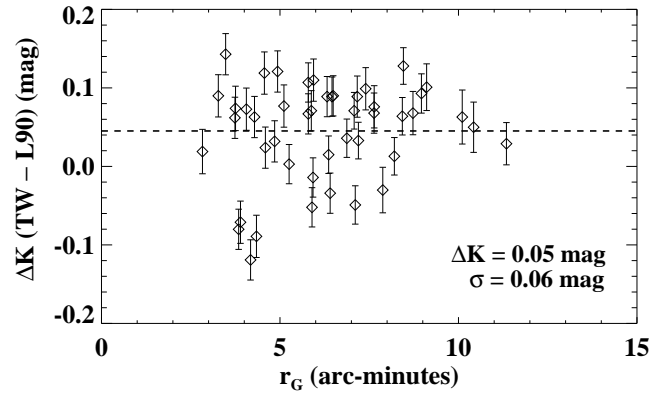}
\caption{Difference in $K$-band mean magnitudes between our photometry and that of \citet{longmore1990} as a function of the radial distance from the cluster center. TW: This work.
The median value and the standard deviation are also shown.} 
\label{fig:delk}
\end{figure}

We compared our mean magnitudes with those from \citet{longmore1990}. The magnitudes from \citet{longmore1990} were in the AAO photometric system. For a relative comparison, the photometric transformations{\footnote{\url{https://www.astro.caltech.edu/~jmc/2mass/v3/transformations/}}} from \citet{carpenter2001} are used to convert AAO magnitudes to the 2MASS system. These transformations also require a $(J-K)_{\textrm{AAO}}$ color term which has a coefficient of $-0.01$~dex. Since $J$-band magnitudes were not provided by \citet{longmore1990}, we adopt a median $(J-K_s)$ color for the RRL in our sample. Given the small coefficient of the $(J-K)_{\textrm{AAO}}$ color term, any deviation from the median value within the RRL color range does not make any significant difference to the $K$-band magnitudes. Fig.~\ref{fig:delk} shows the difference in the $K$-band photometry as a function of the radial distance from the center of the cluster. While several common stars show large offsets ($>0.1$~mag), no statistically significant difference can be determined given the scatter around the median value.

\begin{figure}
\epsscale{1.2}
\plotone{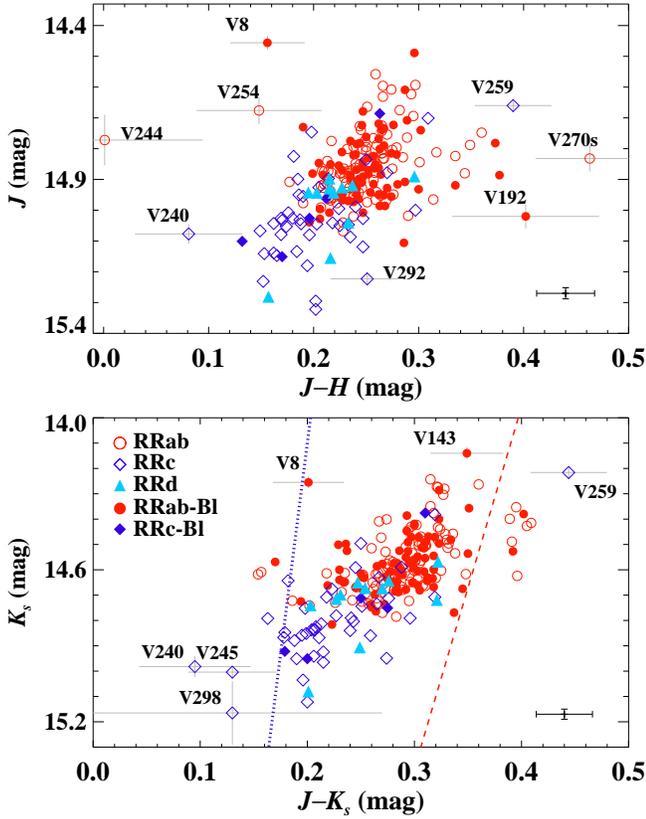}
\caption{NIR color--magnitude diagrams in $(J-H), J$ (top) and $(J-K_s), K_s$ (bottom) for the horizontal-branch RRL. Note that one of the RRL (V297) is not shown 
(see Fig.~\ref{fig:cmd_all}). In the bottom panel, the dotted blue and dashed red lines display the theoretically predicted first overtone blue edge and the fundamental red edge from \citet{marconi2015}.  Some RRL that appear to be located farther from the majority of the variables are marked in each panel and their error bars are also shown. Representative median error bars are also shown at the bottom right of each panel.} 
\label{fig:cmd_jhk}
\end{figure}

\begin{figure*}
\epsscale{1.0}
\plotone{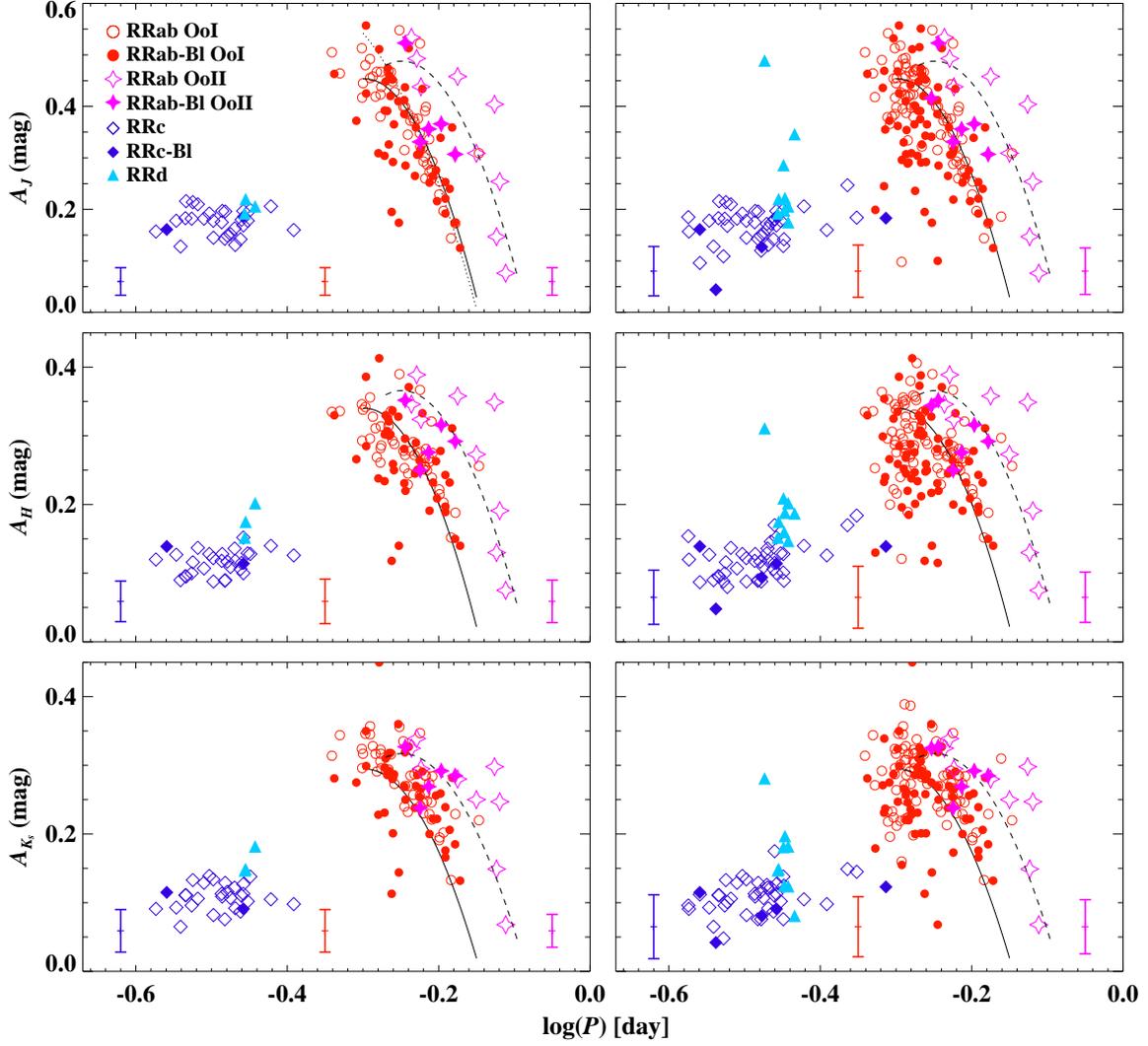}
\caption{{\it Left:} Bailey diagrams for RRL stars in M3 based on good quality (flag `A') light curves in the $J$ (top), $H$ (middle), and $K_s$ (bottom) bands. Overplotted dashed and solid lines represent approximate loci of Oosterhoff I and II types of RRab stars. Dotted line in the top-left panel displays the locus of OoI RRab in the $B$-band from \citet{cacciari2005} scaled arbitrarily by $35\%$. {\it Right:} As the left panels but for RRL with both quality flags `A' and `B'. Representative median error bars are also shown at the bottom of each panel.}
\label{fig:bailey_jhk}
\end{figure*}

\subsection{Color--magnitude and Bailey diagrams}

We used the mean magnitudes and amplitudes estimated from the best-fitting templates to study the pulsation properties of RRL at NIR wavelengths. Fig.~\ref{fig:cmd_jhk} displays
the color--magnitude diagrams in $J-H, J$ and $J-K_s, K_s$ for RRL in M3. The intrinsic color variations in the NIR bands are significantly ($\sim 3-4\times$) smaller 
than in the optical bands. The RRab and RRc pulsators overlap in the so-called ``OR'' region \citep{bono1997b} where both pulsation modes are possible. Most Blazhko RRL 
are also located centrally along the overlapping region between RRab and RRc. In both color--magnitude diagrams, a few RRL that appear to be located farther from the concentrated 
cluster of sources are marked. Most of these exhibit large photometric uncertainties in at least one filter.

In Fig.~\ref{fig:cmd_jhk}, the theoretically predicted fundamental mode red edge and the first-overtone blue edge from \citet{marconi2015} are also overplotted in the $(J-K_s)$, $K_s$ 
color--magnitude diagram. Most NIR observations fall in the region within the predicted boundaries of the instability strip while some RRL are redder/bluer than the fundamental/first-overtone edges. 
Extinction corrections are not applied to the color--magnitude diagrams because the reddening, $E(B-V)=0.01$~mag, in M3 is negligible \citep[][]{harris2010}. Nevertheless, the outlier RRL stars will fall inside the predicted boundaries of the instability strip within $\pm 3\sigma$ of their quoted uncertainties. While the predicted topology of the instability strip may be independent of the metal abundance in the NIR bands, note that the model computations also have a typical minimum resolution of $\pm50$ K in effective temperature \citep{marconi2015}.

Fig.~\ref{fig:bailey_jhk} shows the period--amplitude or Bailey diagrams in the $JHK_s$-bands for M3 RRL variables for the first time. The left panels display Bailey diagrams based on amplitudes determined accurately from the well sampled light curves. In the $J$-band, the amplitudes of the RRab decrease as a function of increasing period similar to the situation in the optical bands \citep[see Fig. 1 of][]{jurcsik2017}. The right panels show Bailey diagrams for light curves with both quality flags `A' and `B'. The loci of OoI and OoII type RRab were determined by fitting second-order polynomials to $J$-band amplitudes in the period range $-0.3 < \log(P) < -0.1$ day, and the following equations were obtained:

\begin{eqnarray}
A_{J_{\textrm{OoI}}} &=& -1.27 (0.09)     -11.59 (0.70) \log(P) \nonumber \\
		     & & -19.47 (1.40) \log(P)^2, \nonumber \\ 
A_{J_{\textrm{OoII}}} &=&  -0.62 (0.16)      -8.93 (1.88) \log(P) \nonumber \\ 
		    & & -17.97 (5.31) \log(P)^2 . 
\end{eqnarray}

The locus of OoI RRab stars is consistent with the scaled optical band locus for RRL in M3 from \citet[][see top left panel of Fig.~\ref{fig:bailey_jhk}]{cacciari2005}. The mean period offset between our empirical OoI and OoII loci is also consistent with the observed shift in the break period in the RRab amplitude ratios in M3 and $\omega$ Cen.  
The $J$-band loci were scaled arbitrarily by 75\% and 65\% in the $H$ and $K_s$-bands to provide a relative comparison of amplitudes with different quality flags. The majority of amplitudes for RRab that were determined from the light curves with quality flag `B' fall below the locus of OoI types. This suggests that the amplitudes for the light curves with large phase gaps are likely underestimated because the NIR-to-optical amplitude ratios, used to constrain the amplitudes, exhibit a scatter of $\sim 20\%$ around the mean values. Furthermore, the $V$-band amplitudes of Blazhko RRL in M3 can change by $\Delta V=0.65$~mag, and exhibit a relative change of up to $90\%$ in total amplitude \citep{jurcsik2017}, and therefore, the amplitude estimates are likely uncertain in these cases.

At NIR wavelengths, \citet{braga2018} found evidence that the locus of RRab stars starts to flatten for longer periods while \citet{gavrilchenko2014} found a nearly flat locus of RRab at mid-infrared wavelengths. In Fig.~\ref{fig:bailey_jhk}, the range of amplitudes for RRab in the $H$ and $K_s$-bands is smaller than in the $J$-band and exhibits more scatter. However, no evidence of flatness is noted. Instead a steady decrease in amplitudes is seen for longer period RRab stars. The light curves of RRc are nearly sinusoidal with smaller variability amplitudes, and therefore, the amplitude are well constrained even for light curves with poor phase coverage. Furthermore, no obvious trend is seen in the amplitudes as a function of the radial distance from the cluster center. For low amplitude RRc stars with $A_V <0.1$~mag, the precision of our photometry is insufficient to detect variability in NIR which has smaller amplitude than in optical bands. RRL variables with known Blazhko modulations \citep{jurcsik2015, jurcsik2017} are also overplotted in Fig.~\ref{fig:bailey_jhk}. No obvious trend is seen between Blazhko and non-Blazhko variables unlike in optical bands where Blazhko stars typically exhibit smaller amplitudes at a given period. This is expected in the NIR where no significant amplitude modulations are seen \citep{jurcsik2018} but observations sampled over a long time interval are needed to notice these long-term variations.

\section{Period--Luminosity relations} 
\label{sec:rrl_plr}

\begin{figure*}
\epsscale{1.2}
\plotone{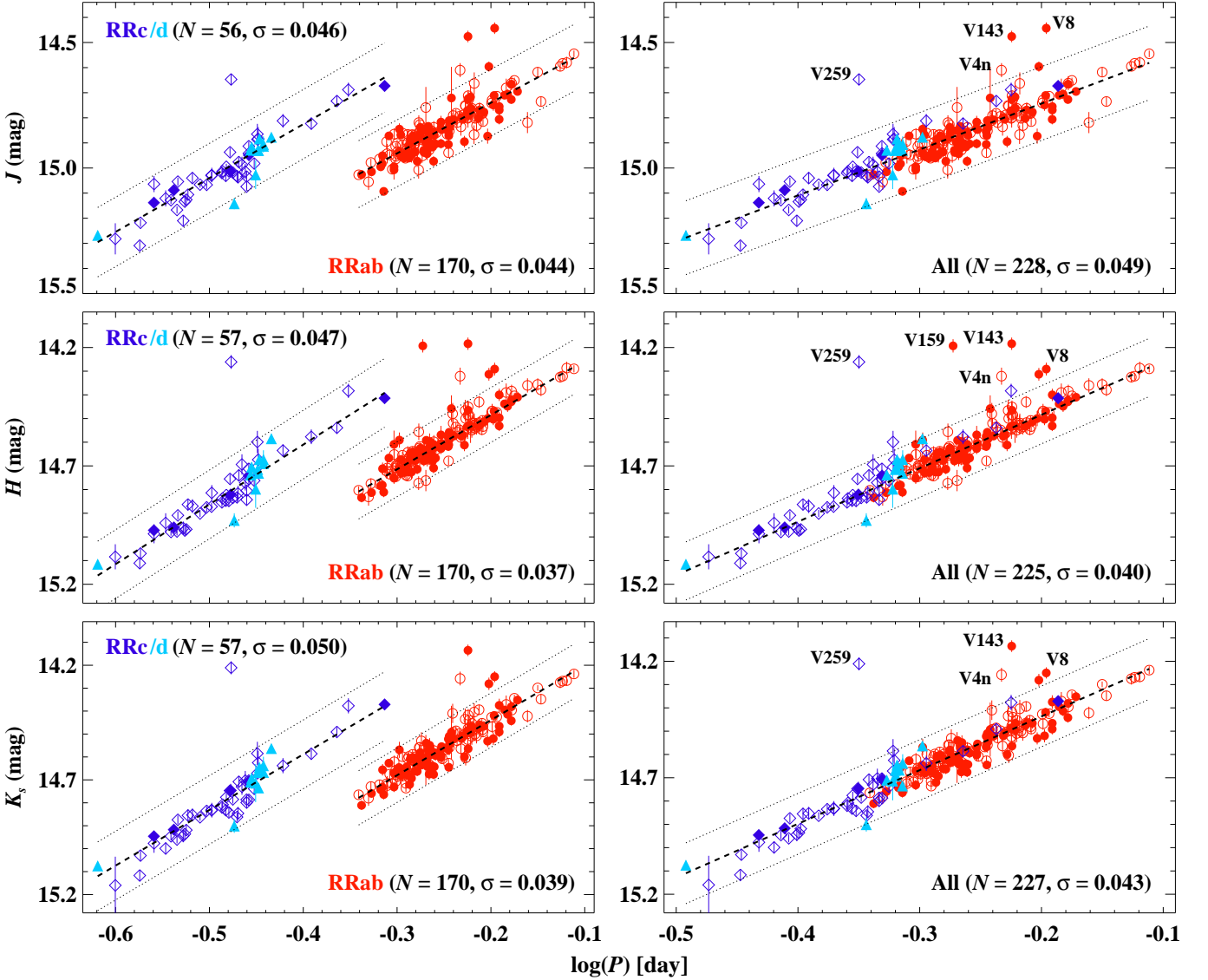}
\caption{NIR PLRs for M3 RRab and RRc+RRd (left) and all RRL (right) in $J$ (top), $H$ (middle), and $K_s$ (bottom) are shown based on our photometry. 
The dashed lines represent best-fitting linear regression over the period range under consideration while the dotted lines display $\pm 3\sigma$ offset from the best-fitting PLRs.
Symbols have the same meaning as in Fig.~\ref{fig:cmd_jhk}. In the right panels, $3\sigma$ outliers with the largest residuals are also marked with the ID of the RRL variable.} 
\label{fig:nir_plr}
\end{figure*}

We used the mean magnitudes listed in Table~\ref{tbl:m3_nir} to derive PLRs for M3 RRL at NIR wavelengths. The reddening in M3 is small - $E(B-V) = 0.01$~mag \citep{harris2010},
0.013 mag \citep{vandenberg2016}. Adopting the reddening law of \citet{card1989} and a total-to-selective absorption ratio $R_V=3.23$, the extinction in the $V$-band amounts to $\sim 0.04$~mag. Therefore, extinction corrections of 13, 9, and 5 mmag were estimated in the $J$, $H$, and $K_s$-bands, respectively, using total-to-selective absorption ratios from \citet{bhardwaj2017a}. 

Under the basic assumption that the PLRs are linear over the entire period range under consideration, the following relation was fitted to the data:
\begin{equation}
{m_\lambda} = a_\lambda + b_\lambda\log(P),
\label{eq:plr}
\end{equation}

\noindent where $a_\lambda$ and $b_\lambda$ give the slope and zero-point of the PLR in a given filter. The scatter ({\it rms}) in the PLR mainly results from the intrinsic width in temperature of the instability strip, a metallicity contribution \citep[$\sim-0.18$~mag/dex in the $K_s$-band,][]{marconi2015} and uncertainties in the extinction correction.  However, the extinction correction uncertainties are minimal in NIR bands and high-resolution spectra of bright stars show that M3 has no appreciable spread in metallicity \citep[$\sigma_{\textrm{[Fe/H]}}=0.03$~dex,][]{sneden2004}.

\begin{deluxetable}{rrrrrr}
\tablecaption{NIR PLRs of RRL in the M3 cluster. \label{tbl:plr_nir}}
\tabletypesize{\footnotesize}
\tablewidth{0pt}
\tablehead{\colhead{Band} & \colhead{Type} & \colhead{$b_\lambda$} & \colhead{$a_\lambda$} & \colhead{$\sigma$}& \colhead{$N$}\\}
\startdata
     $J$ &     RRab & $    14.336\pm0.014     $ & $    -2.018\pm0.052     $ &      0.044 &  170\\
         &    RRc/d & $    13.967\pm0.035     $ & $    -2.145\pm0.073     $ &      0.046 &   56\\
         &      All & $    14.377\pm0.009     $ & $    -1.830\pm0.031     $ &      0.049 &  228\\
     $H$ &     RRab & $    14.027\pm0.016     $ & $    -2.293\pm0.059     $ &      0.037 &  170\\
         &    RRc/d & $    13.601\pm0.042     $ & $    -2.523\pm0.085     $ &      0.047 &   57\\
         &      All & $    14.033\pm0.010     $ & $    -2.258\pm0.034     $ &      0.040 &  225\\
   $K_s$ &     RRab & $    13.959\pm0.015     $ & $    -2.404\pm0.057     $ &      0.039 &  170\\
         &    RRc/d & $    13.618\pm0.045     $ & $    -2.427\pm0.092     $ &      0.050 &   57\\
         &      All & $    13.976\pm0.010     $ & $    -2.305\pm0.035     $ &      0.043 &  227\\
\enddata
\tablecomments{The zero-point ($b$), slope ($a$), dispersion ($\sigma$) and the number of stars ($N$) in the final PLR fits are tabulated.}
\end{deluxetable}

We considered three different samples of RR Lyrae to derive PLRs: (1) RRab variables; (2) a combined sample of RRc and RRd, where dominant first-overtone periods are used for the latter; (3) a combined sample of RRab, RRc, and RRd variables after fundamentalizing overtone periods using $\log(P_{\textrm{FU}})=\log(P_{\textrm{FO}})+0.127$, where `FU' and `FO' represent fundamental and first-overtone modes. Note that 5 RRL with periods shorter than $\sim 0.297$ days are pulsating in the second overtone mode \citep[see Table 1 of][]{jurcsik2015}. 

Fig.~\ref{fig:nir_plr} displays $JHK_s$-band magnitudes for the RRL in M3 plotted as a function of the logarithm of their pulsation periods. We fitted a linear regression
in the form of Eq.~(\ref{eq:plr}) iteratively removing the single largest $>3\sigma$ outlier in each filter separately until convergence. The best-fitting PLRs are also shown in Fig.~\ref{fig:nir_plr} and the results of the regression analysis are listed in Table~\ref{tbl:plr_nir}. The scatter in the empirical $JHK_s$-band PLRs is consistently $\lesssim 0.05$~mag which is up to twice smaller than that in the optical $RI$-band PLRs. Adopting a smaller sigma-clipping threshold ($\sim 2\sigma$), the scatter in these relations is only limited to the photometric uncertainties while allowing us to retain $\sim 75\%$ of RR Lyrae within this threshold. We also investigated possible variations in the slopes and zero-points of the PLRs for samples with light curve quality flags `A' and `B', and found no statistically significant differences from the values quoted in Table~\ref{tbl:plr_nir}. Furthermore, we also found consistent results in terms of the slopes and zero-points of the PLRs after excluding:  (1) Blazhko stars, (2) second-overtone mode variables, (3) stars within $1.5'$ radius from the center of the cluster, (4) stars within a period bin of $\log(P) = 0.05$ day at either end of the period distribution under consideration. 

\begin{figure}
\epsscale{1.2}
\plotone{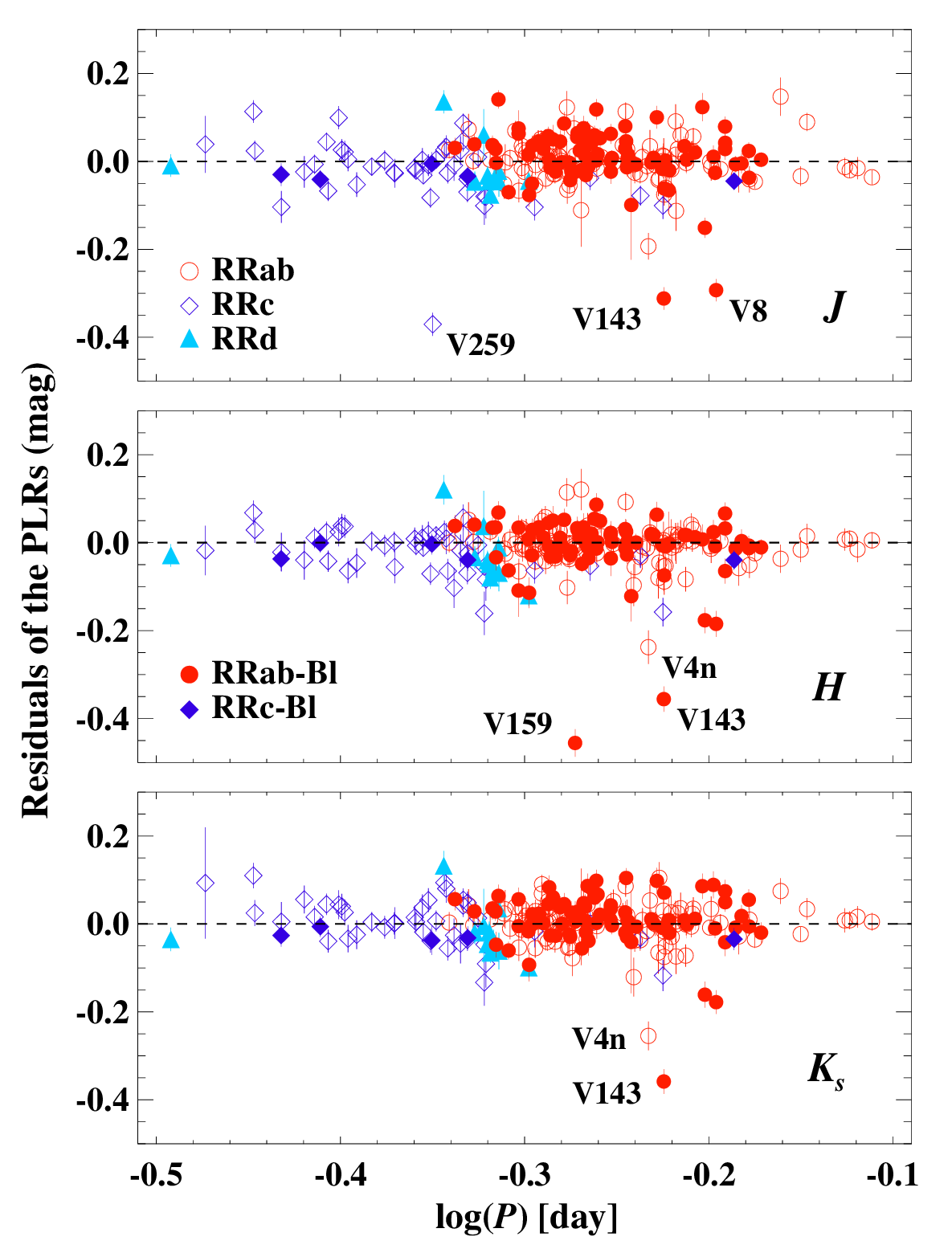}
\caption{Residuals of the PLR fits to the combined sample of RRL stars in the $J$ (top), $H$ (middle), and $K_s$-band (bottom) plotted against the logarithm of the pulsation period. V259 is located outside the {\it y}-axis range shown in the middle and bottom panels.} 
\label{fig:res_per}
\end{figure}

Fig.~\ref{fig:res_per} shows the residuals of the $JHK_s$-band PLR fits plotted against the logarithm of the pulsation period. We do not observe any distinct trend in the residuals of the PLRs except that the majority of RRab stars with periods close of 0.5 days exhibit positive residuals in the $J$-band. On the other hand, the majority of RRc stars in the overlapping period range seem to exhibit negative residuals. Note that RRab stars with periods close to 0.5 days also exhibit large phase gaps due to the observing cadence. This can lead to an offset in their mean magnitudes if the amplitudes of the template fits are not well constrained. The residuals of the PLRs for RRL, which are located in the central $1.5'$, are also consistent with zero-mean. However, these residuals exhibit standard deviations ($\sim$0.09 mag) up to two times larger than for those in the outer regions. Furthermore, some of the outliers with the largest residuals (including V143) are also known to exhibit Blazhko effects. A discussion about individual RRL including outliers in the PLRs is given in the Appendix \ref{sec:app_a}. 

We also compared the residuals of the PLRs against the spectroscopic metallicities provided by \citet{sandstrom2001} for 27 RRab variables in common. \citet{sandstrom2001} determined metallicities using iron lines from moderate-resolution spectra and found a mean [Fe/H]$_{\textrm{FeI}}=-1.22$~dex with a standard deviation of 0.12 dex. However, the median uncertainties in their measurements are of the order of $0.15$~dex and the metallicity range is minimal ($\Delta$[Fe/H]$\sim$0.36~dex) given the uncertainties. We do not observe any obvious trend in the residuals against the metallicity which is expected as high-resolution spectra of bright giants do not provide any evidence of a significant spread in the mean metallicity of M3 \citep{sneden2004}.

\begin{figure}
\epsscale{1.2}
\plotone{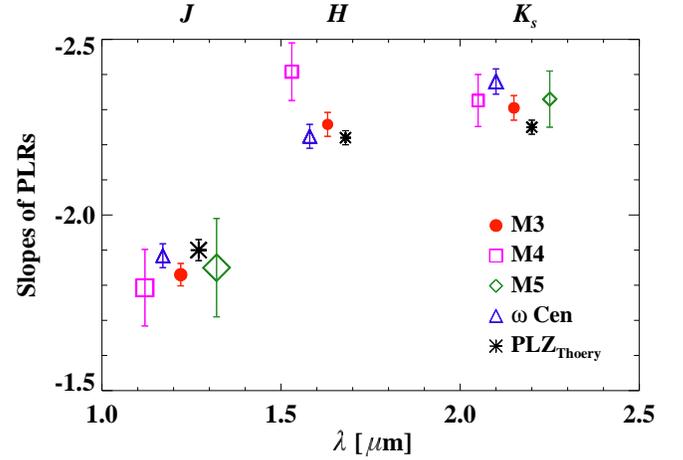}
\caption{Comparison of the slopes of PLRs of RRL in GCs in the $J$, $H$, and $K_s$-bands. The data points corresponding to the slopes from different studies in the literature are slightly offset along the {\it x}-axis for visual clarity. A larger symbol size represents a larger dispersion in the underlying PLRs. The slopes in different GCs are adopted from: M4 \citep{braga2015}, M5 \citep{coppola2011}, $\omega$ Cen \citep{braga2018}, and theoretical results were taken from \citet{marconi2015}. }
\label{fig:slp_plr}
\end{figure}

Finally, we also compared the slopes of the NIR PLRs of RRL in GCs as shown in Fig.~\ref{fig:slp_plr}. A well-known trend in the slopes of NIR PLRs, which become steeper when moving from the $J$ to $K_s$ bands  \citep{neeley2017, beaton2018, bhardwaj2020}, is also seen for M3 RRL variables. The slopes of the $JHK_s$-band PLRs are consistent with those for RRL in the GCs with different mean-metallicities \citep[{[Fe/H]}$=-1.50$ in M3; $-1.16$ in M4, $-1.29$ in M5,][]{harris2010}, and in the GC with a significant spread in metallicity \citep[$\omega$ Cen,][]{braga2018}. Furthermore, our PLR slopes for samples of RRab and all RRL are in good agreement with theoretically predicted PLZ relations \citep[$J$: $-$1.98 (RRab), $-$1.90 (all); $H$: $-$2.24 (RRab), $-$2.22 (all); $K_s$: $-$2.27 (RRab), $-$2.25 (all),][]{marconi2015}. The PLR slopes for RRc stars are shallower than the theoretical predictions in all three bands but statistically consistent given the larger uncertainties.

\subsection{Distance to the M3 cluster}

\begin{deluxetable*}{r c c c c c c c c c}
\tablecaption{Distance to the M3 cluster. \label{tbl:mu_m3}}
\tabletypesize{\footnotesize}
\tablehead{ \colhead{} & \multicolumn{3}{c}{$J$} & \multicolumn{3}{c}{$H$} & \multicolumn{3}{c}{$K_s$}\\
\cline{2-4} \cline{5-7} \cline{8-10}
\colhead{RRL} &$\mu$ &$\sigma_{\textrm{stat.}}$ &$\sigma_{\textrm{syst.}}$ &$\mu$ &$\sigma_{\textrm{stat.}}$ &$\sigma_{\textrm{syst.}}$ &$\mu$ &$\sigma_{\textrm{stat.}}$ &$\sigma_{\textrm{ syst.}}$ \\
& mag	& mag	& mag	& mag	& mag	& mag	& mag	& mag	& mag	}
\startdata
&\multicolumn{9}{c}{Theoretical PLZ calibrations from \citet{marconi2015}}	\\
\hline
   RRc/d &    15.083 &     0.050 &     0.057 &     15.031 &     0.032 &      0.058 &     15.036 &     0.031 &      0.058\\ 
    RRab &    15.077 &     0.020 &     0.037 &     15.046 &     0.017 &      0.033 &     15.046 &     0.017 &      0.037 \\ 
     All &    15.082 &     0.019 &     0.042 &     15.034 &     0.017 &      0.035 &     15.050 &     0.018 &      0.040 \\ 
&\multicolumn{9}{c}{Mean distance =     15.041$\pm$     0.017 (stat.) $\pm$     0.036 (syst.) mag}	\\
\hline
&\multicolumn{9}{c}{Empirical calibrations with {\it HST} parallaxes from \citet{benedict2011}}	\\
\hline
     All\tablenotemark{\small a} &    15.138 &     0.087 &     0.122 &     15.121 &     0.084 &     0.089 &     15.113 &     0.082 &     0.081 \\ 
     All\tablenotemark{\small b} &    15.068 &     0.072 &     0.132 &     15.040 &     0.069 &     0.093 &     15.032 &     0.065 &     0.092 \\
\hline
&\multicolumn{9}{c}{Empirical calibration with {\it Gaia} parallaxes from \citet{muraveva2018}} \\
\hline
     All\tablenotemark{\small c} &      ---  &      ---  &      ---  &       ---  &      ---  &       ---  &     15.001 &     0.098 &    0.121\\ 
\enddata
\tablenotetext{\tiny a}{Average distance modulus in $JHK_s$-bands estimated using the zero-point calibration based on individual RRL with {\it HST} parallax. No [Fe/H] correction applied.}
\tablenotetext{\tiny b}{Same as above but with [Fe/H] correction applied.}
\tablenotetext{\tiny c}{Calibration based on the PL$Z_{K_s}$ relation listed in Table 4 of \citet{muraveva2018}.}
\end{deluxetable*}

New NIR photometry of RRL in M3 provides an opportunity to estimate a robust distance to the cluster thanks to the precision and accuracy of the mean magnitudes and derived PLRs. However, an absolute calibration of NIR PLRs of RRL is still lacking, and the precision of the estimated distances is mainly affected by the zero-point uncertainties of the calibrator relations \citep{beaton2018, muraveva2018, bhardwaj2020}. Theoretical models predict a significant metallicity dependence of the NIR PLRs \citep{catelan2004, marconi2015} but
some empirical relations also suggest a marginal or weaker dependence on metallicity \citep{sollima2006, muraveva2015}. Note that theoretical calibration has been preferred in the most recent studies on distance determination using infrared observations of RRL \citep[e.g.,][]{neeley2017, braga2018}. 

First, we also adopted the theoretical calibrations of the RRL PLZ relation in the $JHK_s$-bands from \citet{marconi2015}. Given that the metallicities in these predicted relations are on the \citet{carretta1997} scale, an iron-abundance of [Fe/H]$=-1.34$~dex is adopted for M3. \citet{marconi2007a} also modeled the light curves of RRL in M3 for [Fe/H]$=-1.3$ dex, which led to a metal-abundance $Z\sim$ 0.001, and estimated a distance modulus to the cluster, $\mu=15.10 \pm 0.10$~mag. The slope and metallicity coefficients of the predicted relations were used to anchor the absolute zero-point for the $JHK_s$-band PLRs and determine a distance modulus to the M3 cluster. The results of distance measurements using $JHK_s$-band PLRs are tabulated in Table~\ref{tbl:mu_m3}. Distance moduli based on $J$-band PLZ relations are comparatively larger than for the $H$ and $K_s$-bands possibly due to differences in the slopes and a relatively larger dispersion in the calibrator relations ($0.06$~mag for RRab and All). The statistical uncertainties were quantified through propagation of the errors in the photometry and uncertainties in the coefficients of the predicted relations. For systematic uncertainties, errors in the zero-points, errors in the slopes propagated through the difference of the mean periods between calibrator and cluster PLRs, and uncertainties due to possible mean metallicity variations ($\Delta$[Fe/H]=0.05~dex) were added in quadrature. Using the weighted mean of the $H$ and $K_s$-band measurements, we determined a distance to the M3 cluster of $\mu = 15.041 \pm 0.017~(\textrm{stat.}) \pm 0.036~(\textrm{syst.})$ mag.

We also employed an empirical calibration based on five Galactic RRL with trigonometric parallaxes from the {\it Hubble Space Telescope (HST)} Fine Guidance Sensor \citep{benedict2011}. 
NIR mean magnitudes for these calibrator RRL were adopted from \citet{feast2008} and \citet{monson2017}. The small sample of RRL and their modest period range ($-0.51 \gtrsim \log(P) \lesssim -0.18$ day) do not allow for good constraints on the slopes and zero-points of the PLRs. Therefore, absolute zero-points of the PLRs listed in Table~\ref{tbl:plr_nir} were determined based on the {\it HST} parallaxes of individual RRL variables. A weighted mean of the distances to M3 based on five RRL was adopted as the distance modulus to the cluster and the results are given in Table~\ref{tbl:mu_m3}. Empirical calibrations of PLRs based on {\it HST} parallaxes typically lead to a larger distance modulus to M3. This is expected since no metallicity term is included in the PLRs in Table~\ref{tbl:plr_nir}, and on average the 5 Galactic RRL with the HST parallaxes are more metal-poor \citep[{[Fe/H]}$\sim-1.63$~dex,][see Table 8]{bhardwaj2016a} than the mean metallicity of M3. Accounting for the metallicity term according to the predicted PLZ relations, the distance measurements based on empirical relations also become consistent with the value obtained using theoretical calibrations. However, 
\citet{neeley2017} suggested that the {\it HST} parallaxes for the calibrator RRL and their reddening values in the literature may be affected by systematics, in particular for RR Lyr and UV Oct. Indeed, the parallax of RR Lyr yields the largest distance modulus to M3 despite having [Fe/H]$=-1.39$~dex, which is more consistent with M3.

We also used an empirical calibration of the PLZ$_{K_s}$ relation based on {\it Gaia} parallaxes \citep{muraveva2018} and found a distance estimate consistent with those based on theoretical calibration. However, uncertainties in the distance estimates based on {\it Gaia} parallaxes are large given a systematic zero-point offset present in the current data release \citep[see,][for details]{muraveva2018}.
From Table~\ref{tbl:mu_m3}, it is evident that the distances determined using the predicted $JHK_s$-band PLZ relations have the smallest uncertainties for the sample of RRab and the combined RRL sample. As mentioned, the empirical PLRs for RRc stars are shallower than the predicted relations, and the latter also exhibit relatively larger errors in the calibrator coefficients. The agreement between three RRL samples in three different filters is within $1\sigma$ of the quoted systematic uncertainties. Given the larger systematics in both {\it HST} and {\it Gaia} DR2 parallaxes, the distance modulus based on the theoretical calibration is adopted to estimate a distance $D = 10.19 \pm 0.08~(\textrm{stat.}) \pm 0.17~(\textrm{syst.})$ kpc to M3.

Recent estimates of the distance modulus to M3 range between $15$ and $15.1$~mag based on several independent methods, for example, a value of 15.07~mag is quoted in the GC catalog of \citet{harris2010}. Using an empirical PLZ$_{K_s}$ relation, \citet{sollima2006} estimated a distance modulus of 15.07~mag to M3. \citet{vandenberg2016} used a distance modulus of $15.04$~mag to perfectly fit observations using zero-age horizontal branch evolutionary models and \citet{tailo2019} found a value of 15.07 mag using main-sequence isochrone fitting to the cluster color--magnitude diagrams. Based on the Baade--Wesselink method, \citet{jurcsik2017} determined a distance modulus of 15.10$\pm$0.043~mag to the M3 cluster. Our final distance modulus, $\mu = 15.041 \pm 0.017~(\textrm{stat.}) \pm 0.036~(\textrm{syst.})$ mag, is in excellent agreement with the independent M3 distance estimates in the literature.

\section{Summary} \label{sec:discuss}

We have presented new NIR time-series observations of a $21'\times 21'$ sky area around the center of the M3 globular cluster. Our sample of RRL in M3 was adopted from the catalog of \citet{clement2001}, and uses accurate pulsation periods and $V$-band amplitudes from the extensive optical photometric studies in the literature \citep[for example, ][]{jurcsik2015, jurcsik2017}. The ensemble NIR photometry from multi-epoch observations was derived and calibrated with an internal photometric precision of better than $2\%$ for RRL in moderately crowded regions. Combining optical and NIR data resulted in the largest sample to date of 233 RRL in a single cluster with multi-epoch $JHK_s$-band data. We used light curve data to investigate amplitude ratios and Bailey diagrams for RRL in the $JHK_s$-bands for the first time in M3. New templates for RRL in NIR from \citet{braga2018} were used to determine precise photometric mean magnitudes in the $JHK_s$-bands and derive new PLRs. Our precise PLRs will be useful to investigate the dependence on metallicity when complemented with literature data of homogeneous RRL populations in the GCs having independent distances and different mean-metallicities.

We summarize our main results as follows :

\begin{itemize}
\item{We presented $JHK_s$-band light-curve data for 233 RRL variables in the M3 cluster with an average of 20 epochs in each filter. The M3 RRL sample consists of 175 RRab, 47 RRc and 11 RRd variables with $JH$-band time-series for the first time. It also provides a five-fold increase in the sample size of M3 RRL with $K_s$-band mean magnitudes available in the literature.}

\item{NIR-to-optical amplitude ratios for RR Lyrae in M3 display a systematic increase moving from RRc to short-period ($P < 0.6$~days) and long-period ($P < 0.6$~days) RRab variables. Similar trend is also observed in the amplitude ratios ($A_{HK_s}/A_J$) involving only NIR bands.  The shift in the median values of the amplitude ratios for long-period RRab occurs at an earlier period for M3 variables than for those in the $\omega$ Cen. This observed shift in the break period ($\Delta \log(P) = -0.067$ [days]) is in excellent agreement with the difference between the mean RRab periods in the two distinct Oosterhoff type clusters (OoI M3 and OoII $\omega$ Cen).} 

\item{The largest sample of RRL (or RRab) in a single cluster is used to derive new $JHK_s$-band PLRs. Our sample of 175 RRab stars encompasses almost twice the number of fundamental pulsators in $\omega$  Cen with time-series NIR photometry \citep{braga2018}. The residuals of these empirical relations do not display any trend as a function of metallicity
suggesting that the spread in metallicity of individual M3 RRL is negligible.}

\item{The slopes of empirical $JHK_s$-band PLRs for M3 RRL are in excellent agreement with the slopes of PLRs for RRL in GCs with different mean-metallicities. Furthermore, our PLRs for RRab and the combined sample are also consistent with the theoretical predictions of the PLZ relations from \citet{marconi2015}. While PLRs for RRc stars are shallower than the theoretical predictions, they are also in agreement within the uncertainties.}

\item{We used predicted RRL PLZ relations with a chemical abundance of $Z=0.001$, $Y=0.25$, to determine a distance modulus to M3 of $\mu = 15.041 \pm 0.017~(\textrm{stat.}) \pm 0.036~(\textrm{syst.})$~mag. Our distance estimate is in a very good agreement with distances determined based on modeling of M3 RRL \citep{marconi2007a} or the Baade--Wesselink method \citep{jurcsik2017}. We also found consistent distance estimates based on the zero-point calibration using {\it HST} or {\it Gaia} parallaxes for RRL provided a proper account of metallicity effects is taken.}
\end{itemize}

\acknowledgements

We thank the anonymous referee for the quick and constructive referee report that helped improve the manuscript.
We also thank Lucas Macri, Tarini Konchady, and Peter B. Stetson for kindly answering queries related to the data reduction and photometric analyses. 
We are also thankful to Eric Peng, Laurie Rousseau-Nepton, and Pascal Fouqu\'e for general discussions on scheduling observations and pre-processing of WIRCam data, 
and Vittorio F. Braga for reading an earlier version of the manuscript. 
AB acknowledges research grant $\#11850410434$ from the National Natural Science Foundation of China through the Research Fund for International Young Scientists,
a China Post-doctoral General Grant, and the Gruber fellowship 2020 grant sponsored by the Gruber Foundation and the International Astronomical Union. HPS and SMK acknowledge the support from the Indo-US Science and Technology Forum, New Delhi, India. CCN is grateful for the funding from Ministry of Science and Technology (Taiwan) under contract 107-2119-M-008-014-MY2. This research was supported by the Munich Institute for Astro- and Particle Physics (MIAPP) of the DFG cluster of excellence ``Origin and Structure of the Universe''. This research uses data obtained through the Telescope Access Program (TAP) of China.\\

\facility{CFHT (WIRCam Near-infrared imager)} 

\software{\texttt{IRAF} \citep{tody1986, tody1993}, \texttt{DAOPHOT/ALLSTAR} \citep{stetson1987} \texttt{DAOMATCH and DAOMASTER} \citep{stetson1993}, \texttt{ALLFRAME} \citep{stetson1994}, \texttt{SExtractor} \citep{bertin1996}, \texttt{SWARP} \citep{bertin2002}, \texttt{SCAMP} \citep{bertin2006}, \texttt{WeightWatcher} \citep{marmo2008}, \texttt{IDL} \citep{landsman1993}, \texttt{Astropy} \citep{astropy2013}}\\\\\\\\

\newpage 

\appendix

\section{Additional figures}
\label{sec:app_b}
NIR light curves of a few randomly selected M3 RRL with different quality flags are shown in Figs.~\ref{fig:lc_aqf} and \ref{fig:lc_bqf}. Time-series data is available online as supplementary material for RRL in M3.

\section{Comments on a few RR Lyrae variables}
\label{sec:app_a}

{\it V4s and V4n:} These are two RRL with similar periods that are separated by 0.45$\arcsec$ and good-quality light curves for both variables were obtained. V4n is significantly brighter than the best-fitting PLRs. Since the $JHK_s$ amplitudes of V4n are up to $23\%$ smaller than those for V4s, it is likely that photometry of this RRL is biased by a few epochs obtained in relatively poorer seeing due to blending with nearby stars.

{\it V8, V159, V259:} These RRL are brighter than the best-fitting PLRs in at least one filter. Photometry of these objects is blended due to bright sources in close proximity.
$J$ and $K_s$-band light curves of V8 display clear periodicity but $H$-band photometry is significantly contaminated. V159 is an outlier only in the $H$-band although there it exhibits a more 
periodic light curve than in the $JK_s$-bands. 

{\it V48 and V143:} We recovered a high quality light curve for V143 with full phase coverage despite it being located within the cluster's unresolved central $1.5'$ region, and the cause of this Blazhko RRL being brighter than the PLRs is not clear. Similarly, V48 is also brighter than the PLRs although it is well-resolved and the light curves are well-sampled in the $JHK_s$-bands. 

{\it V129, V217, V234:} We have confirmed the uncertain classification of these variables listed in the catalog of \citet{clement2001}. V129 is an RRc while V217 is an RRab variable. V234 is marked as a candidate field star but its mean magnitudes are consistent in the $JHK_s$ band PLR plane, and therefore, it is likely a cluster member.  

{\it V148, V181, V242, V246, and V261:} These RRL have proper motions beyond $\pm5\sigma$ of their mean values and exhibit (except for V261) residuals that are consistent within $\pm2\sigma$ in all three $JHK_s$ filters.

{\it V192, V244, and V298:} The light curves do not exhibit any periodicity but the weighted mean magnitudes are consistent with the best-fitting $JHK_s$ band PLRs. V244 exhibits large scatter in the time-series and the residuals of the $JH$-band PLRs are also large ($>0.1$~mag). Similarly, individual measurements for V298 also exhibit large photometric uncertainties. 

{\it V220, V251, and V255:} The light curves of these RRL display periodicity despite large scatter and the mean magnitudes are consistent with the best-fitting PLRs.

{\it V265:} We derived a period (0.5284 days) for the first time and confirm that it is an RRL variable. It is classified as RRab and $JHK_s$ band mean magnitudes are consistent with RRL PLRs. Photometry is severely blended for another close companion, V268, preventing us from obtaining any estimate of the pulsation period.

{\it V297:} This is an obvious outlier in the proper motions, color--magnitude diagrams, and the Period--Luminosity planes. We also looked at the 2MASS magnitudes for this object and found that it is more than a magnitude brighter in the $JHK_s$-bands than the horizontal branch RRL with similar periods. V297 is also significantly redder ($B-V=0.97$~mag) than horizontal branch stars in the optical color--magnitude diagram \citep[][their Figure 8 and Table 2]{hartman2005}. Since the $V$-band amplitude of V297 is very small \citep[$0.05$~mag,][]{hartman2005}, no periodic variability is recovered in our photometry. It is located in the outskirts of the cluster and is unlikely blended, suggesting it is either misclassified as an RRL or it may be a field variable. Therefore, we do not consider V297 a member of the cluster RRL population.

\begin{figure*}
\epsscale{1.2}
\plotone{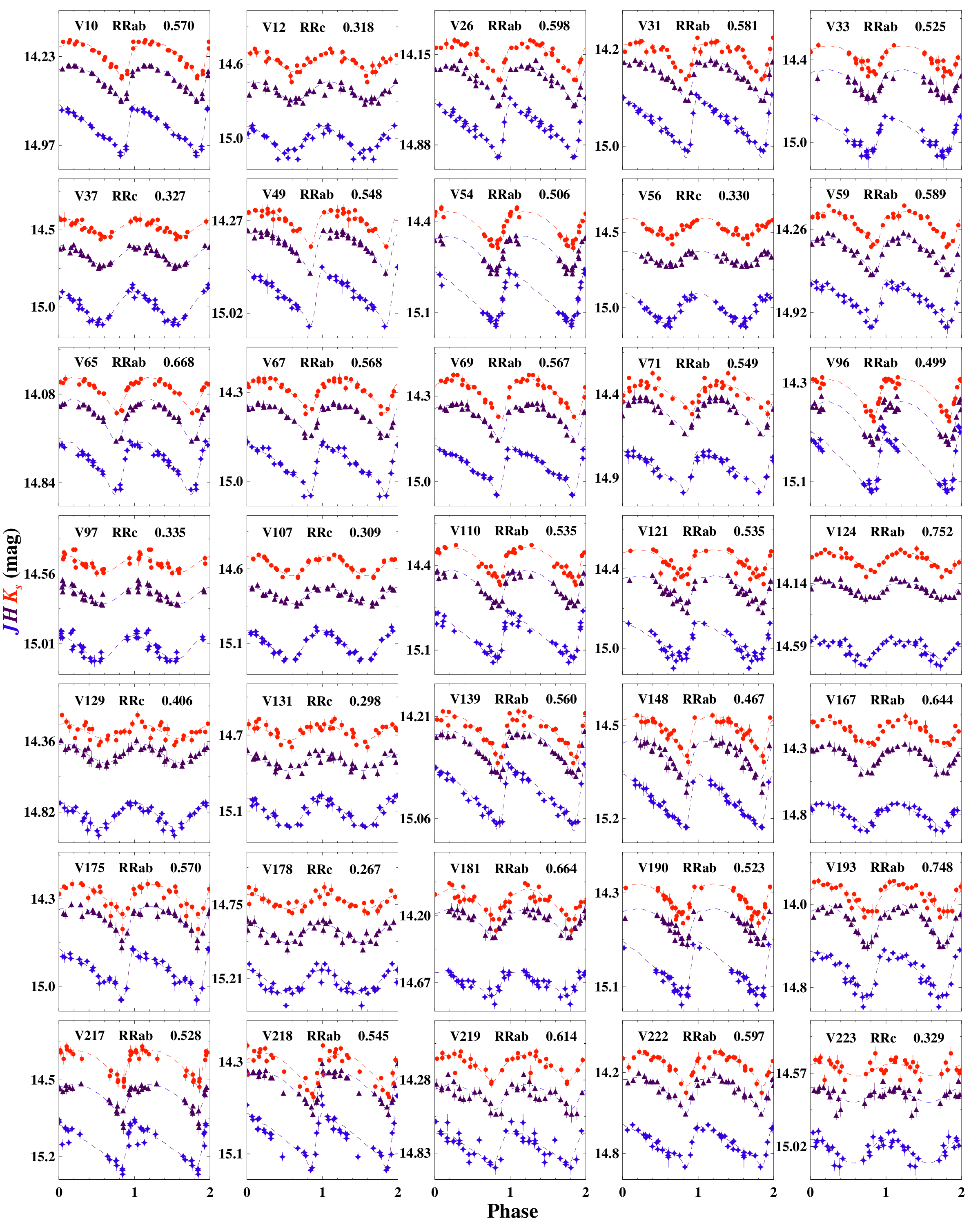}
\caption{Example $JHK_s$-band light curves of RRL with quality flag `A' in our sample. The $J$ (blue stars) and $K_s$ (red circles) light curves are offset for clarity by $+0.1$ and $-0.2$ mag, respectively. The dashed lines represent the best-fitting templates to the data in each band. Star ID, subtype, and the pulsation period are included at the top of each panel.} 
\label{fig:lc_aqf}
\end{figure*}

\begin{figure*}
\epsscale{1.2}
\plotone{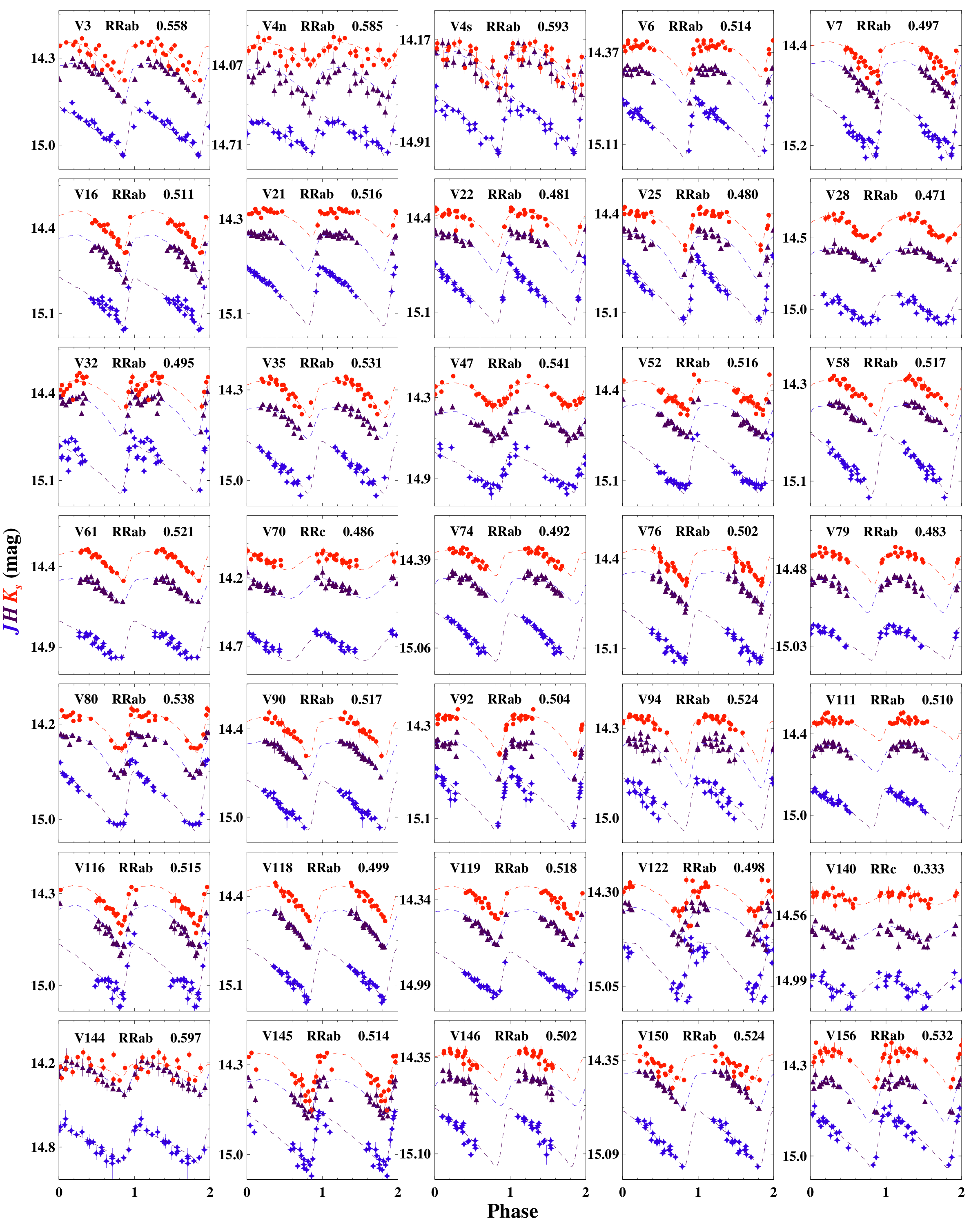}
\caption{As Fig.~\ref{fig:lc_aqf} but for the RRL with light curve quality flag `B'.} 
\label{fig:lc_bqf}
\end{figure*}

\bibliographystyle{aasjournal}
\bibliography{/home/anupam/work/manuscripts/mybib_final.bib}
\end{document}